\documentclass[sigconf]{acmart}
\renewcommand\footnotetextcopyrightpermission[1]{} % removes footnote with conference information in first column
\setcopyright{none}

\usepackage{caption}
\usepackage{subfig}
\usepackage{graphicx}
\usepackage{booktabs}
\usepackage{epstopdf}
\usepackage{url}
\usepackage{placeins}
\usepackage{amsmath,amsfonts}
\usepackage{mathtools,mathrsfs}
\usepackage{multirow}
\usepackage{rotating}
\usepackage{pbox}
\usepackage[normalem]{ulem}
\usepackage{listings}

\usepackage{cleveref}
\usepackage[utf8]{inputenc}
\crefname{section}{§}{§§}
\Crefname{section}{§}{§§}

\usepackage[10pt]{moresize}

\usepackage{xcolor}
\definecolor{darkgrey}{RGB}{70,70,70}
\definecolor{lightgrey}{RGB}{200,200,200}

\usepackage{enumitem}

\usepackage[linesnumbered,ruled,vlined]{algorithm2e}
\newcommand{\goal}[1]{\noindent\textcolor{red}{[Goal: #1]}\par}

\newcommand{\macb}[1]{\textbf{\textsf{#1}}}

\usepackage[hang,flushmargin]{footmisc}

\usepackage{bm}

\usepackage{scalerel,stackengine}
\stackMath
\newcommand\rwh[1]{%
\savestack{\tmpbox}{\stretchto{%
  \scaleto{%
      \scalerel*[\widthof{\ensuremath{#1}}]{\kern-.6pt\bigwedge\kern-.6pt}%
          {\rule[-\textheight/2]{1ex}{\textheight}}%WIDTH-LIMITED BIG WEDGE
            }{\textheight}% 
}{0.5ex}}%
\stackon[1pt]{#1}{\tmpbox}%
}

\def\HiLiGA{\leavevmode\rlap{\hbox to \hsize{\color{black!10}\leaders\hrule height 1\baselineskip depth 1ex\hfill}}}
\def\HiLiGB{\leavevmode\rlap{\hbox to \hsize{\color{black!25}\leaders\hrule height 1\baselineskip depth 1ex\hfill}}}
\def\HiLiGC{\leavevmode\rlap{\hbox to \hsize{\color{black!40}\leaders\hrule height 1\baselineskip depth 1ex\hfill}}}
\def\HiLiGD{\leavevmode\rlap{\hbox to \hsize{\color{black!55}\leaders\hrule height 1\baselineskip depth 1ex\hfill}}}
\def\HiLiGE{\leavevmode\rlap{\hbox to \hsize{\color{black!70}\leaders\hrule height 1\baselineskip depth 1ex\hfill}}}
\def\HiLiGF{\leavevmode\rlap{\hbox to \hsize{\color{black!85}\leaders\hrule height 1\baselineskip depth 1ex\hfill}}}

\usepackage{algpseudocode}
\usepackage{tikz}
\usetikzlibrary{calc}
\usepackage{xcolor}
\makeatletter

\renewcommand{\goal}[1]{}

\begin{document}

\setlength{\pdfpageheight}{\paperheight}
\setlength{\pdfpagewidth}{\paperwidth}

\title{High-Performance Distributed RMA Locks}

\author{Patrick Schmid$^*$, Maciej Besta$^{*\dagger}$, Torsten Hoefler}
       \affiliation{\vspace{0.3em}Department of Computer Science, ETH Zurich\\
       {$^*$}Both authors contributed equally to this work\\
       {$^\dagger$}Corresponding author\\
%       \affaddr{1932 Wallamaloo Lane}\\
%       \affaddr{Wallamaloo, New Zealand}\\
%       \email{trovato@corporation.com}
}

\begin{abstract}
%
% Motivation/problem statement: Why do we care about the problem? What
% practical, scientific, theoretical or artistic gap is your research filling?
%
%We propose what is, to the best of our knowledge, the first topology-aware
%distributed Reader-Writer lock for supercomputers and data centers. 
We propose a topology-aware distributed Reader-Writer lock that
accelerates irregular workloads for supercomputers and data centers.
%
% that tackle two well-known problems related to such settings: expensive
% inter-node data transfers and weak memory models that increase design
% complexity.
%
% the former impacts performance while the latter entails additional design
% challenges.
%
% Methods/procedure/approach: What did you actually do to get your results?
% (e.g. analyzed 3 novels, completed a series of 5 oil paintings, interviewed
% 17 students)
%
The core idea behind the lock is a modular design that is an interplay of three
distributed data structures: a counter of readers/writers in the critical section,
a set of queues for ordering writers waiting for the lock, and a tree that
binds all the queues and synchronizes writers with readers. Each structure is
associated with a parameter for favoring either readers or writers, enabling
adjustable performance that can be viewed as a point in a three
dimensional parameter space. We also develop a distributed topology-aware MCS lock that is a
building block of the above design and improves state-of-the-art MPI
implementations. 
%
% lock with a set of queues for writers that maximize locality and a
% distributed counter for readers for a higher amount of parallelism.  
%
Both schemes use non-blocking Remote Memory Access (RMA) techniques for highest
performance and scalability.
%
% Results/findings/product: As a result of completing the above procedure, what
% did you learn/invent/create?
%
We evaluate our schemes on a Cray XC30 and illustrate that they outperform
state-of-the-art MPI-3 RMA locking protocols by 81\% and 73\%, respectively.
Finally, we use them to accelerate a distributed hashtable that represents
irregular workloads such as key-value stores or graph processing. 
%
% Conclusion/implications: What are the larger implications of your findings,
% especially for the problem/gap identified in step 1?
%
% Our locks can be used to improve highly parallel workloads in today's and
% tomorrow's distributed large-scale processing.
%
%
%
% Inter-node data transfers in distributed-memory machines significantly differ
% from intra-node communication: they of- fer no data coherence and can be many
% times more expen- sive in latency and power consumption. Thus, such ma-
% chines require novel topology-aware synchronization proto- cols for high
% performance parallel codes. In this work, we first develop an MCS lock that
% reduces the number of ex- pensive inter-node lock passings. Second, we
% illustrate a design of a topology-aware distributed Reader-Writer (RW) lock
% with a global queue for writers to maximize locality and a distributed
% counter for readers to trade lock fairness for a higher amount of concurrency
% between readers result- ing in higher throughput. We use state-of-the-art
% Remote Memory Access (RMA) programming techniques for high- est performance
% and scalability. We evaluate our MCS and RW schemes on a Cray XC30 and
% illustrate that they accel- erate state-of-the-art MPI-3 RMA locking
% protocols by ??x and ??x, respectively. Thus, our locks can be used to im-
% prove deeply parallel workloads in today’s and tomorrow’s large-scale
% processing.
%
\end{abstract}

\ccsdesc[500]{Hardware~Testing with distributed and parallel systems}
\ccsdesc[500]{Computer systems organization~Architectures}
\ccsdesc[500]{Computer systems organization~Parallel architectures}
\ccsdesc[500]{Computer systems organization~Multicore architectures}
\ccsdesc[300]{Computing methodologies~Parallel computing methodologies}
\ccsdesc[300]{Computing methodologies~Shared memory algorithms}
\ccsdesc[100]{Hardware~Networking hardware}
\ccsdesc[300]{Computer systems organization~Distributed architectures}
\ccsdesc[100]{Computer systems organization~Multicore architectures}
\ccsdesc[500]{Software and its engineering~Synchronization}
\ccsdesc[300]{Software and its engineering~Access protection}
\ccsdesc[300]{Theory of computation~Parallel algorithms}
\ccsdesc[500]{Theory of computation~Massively parallel algorithms}
\ccsdesc[100]{Theory of computation~Shared memory algorithms}
\ccsdesc[500]{Theory of computation~Distributed algorithms}
\ccsdesc[500]{Computing methodologies~Distributed algorithms}

\maketitle
\pagestyle{plain}

{\vspace{-0.5em}\noindent \textbf{This is an arXiv version of a paper published at\\ ACM HPDC'16 under the same title}}

{\vspace{1em}\small\noindent\textbf{Code:}\\\url{https://spcl.inf.ethz.ch/Research/Parallel\_Programming/RMALocks}}

\section{INTRODUCTION}
\label{sec:intro}

\goal{Explain importance of distributed memory machines for large-scale
processing.}
The scale of today's data processing is growing steadily. For example, the size
of Facebook's social graph is many
petabytes~\cite{bronson2013tao,Venkataramani:2012:TFS:2213836.2213957} and
graphs processed by the well-known HPC benchmark
Graph500~\cite{murphy2010introducing} can have trillions of vertices.
Efficient analyses of such datasets require distributed-memory (DM) machines
with deep \emph{Non-Uniform Memory Access} (NUMA) hierarchies.

\goal{State that locks are important in large-scale processing.}
Locks are among the most effective synchronization mechanisms used in codes for
such machines~\cite{bienia11benchmarking}. On one hand, if used improperly, they may cause
deadlocks. Yet, they have intuitive semantics and they often outperform other
schemes such as atomic operations~\cite{schweizer2015evaluating} or transactions~\cite{besta2015accelerating}.

\goal{Say that deep memory hierarchies are challenging for efficient locks.}
Designing efficient locks for machines with deep hierarchical memory systems is
challenging. Consider four processes competing for the same lock. Assume that
two of them (A and B) run on one socket and the remaining two (C and D) execute
on the other one.  Now, in a naive lock design oblivious to the memory
hierarchy, the lock may be passed between different sockets up to three times,
degrading performance (e.g., if the order of the processes entering the critical section (CS) is
A, C, B, and D).
Recent advances~\cite{Chabbi:2015:HPL:2688500.2688503,
Dice:2012:LCG:2145816.2145848} tackle this problem by reordering processes
acquiring the lock to reduce inter-socket communication. Here, the order of A,
B, C, and D entails only one inter-socket lock transfer, trading fairness
for higher throughput. Extending such schemes to DM machines with weak memory
  models increases complexity. Moreover, expensive inter-node data transfers
  require more aggressive communication-avoidance strategies than those in
  intra-node communication~\cite{fompi-paper}.
%
% Finally, previous NUMA-aware designs cannot be straightforwardly used in DM
% codes.
%
To our best knowledge, no previous lock scheme addresses these challenges.

\begin{figure}[!h]
\centering
\includegraphics[width=0.4\textwidth]{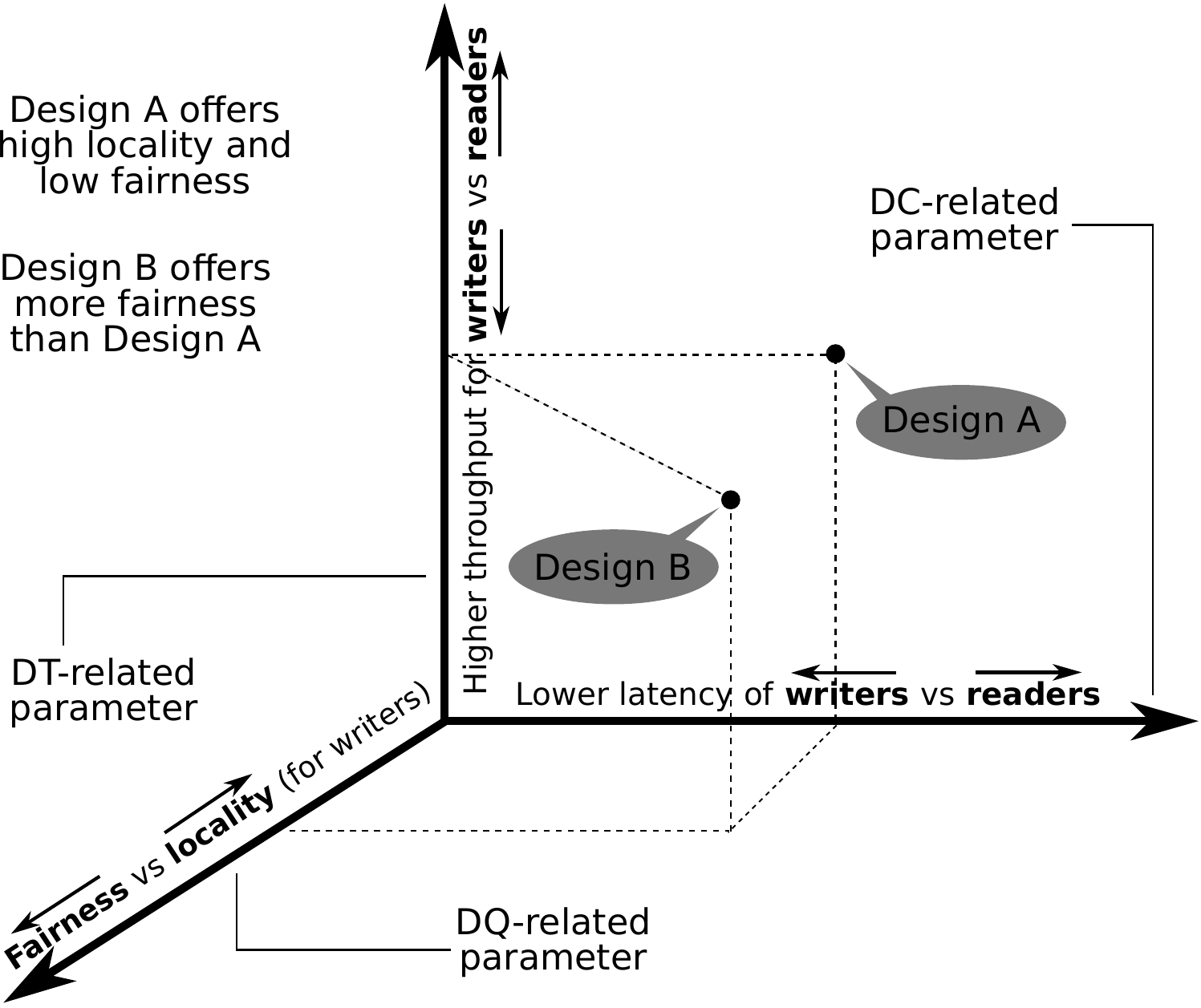}
\caption{The space of parameters of the proposed Reader-Writer lock.}
\label{fig:space}
\end{figure}

% the first
% process would pass the lock to the process on the same socket, then the lock
% would be handed over to a process on the other socket, followed by a final
% intra-socket passing. This gives only one inter-socket lock transfer and it
% trades fairness for higher throughput. Extending such schemes to DM machines
% with weak memory models increases complexity. Moreover, expensive inter-node
% data transfers require more aggressive communication-avoidance strategies than
% those in intra-node communication~\cite{fompi-paper}.
% %
% % Finally, previous NUMA-aware designs cannot be straightforwardly used in DM
% % codes.
% %
% To our best knowledge, no previous lock scheme addresses these challenges.

% \goal{State that we care about RW locks.}
% %
% Developing efficient locks with extended semantics is another challenge in
% today's large-scale processing. The majority of accesses in various analytics
% workloads are reads (e.g., 99.8\% of requests to the Facebook graph are read
% requests~\cite{Venkataramani:2012:TFS:2213836.2213957}). Here, simple locks
% would entail unnecessary synchronization overheads. Instead, the Reader-Writer
% (RW) lock~\cite{Mellor-Crummey:1991:SRS} is a more proper match.  It reduces
% the amount of synchronization for processes that only perform reads in the
% critical section (CS). Initial RW \emph{NUMA-aware} lock designs have recently
% been introduced~\cite{Calciu:2013:NRL}, but they do not address DM machines.

\goal{State that we care about RW locks.}
Another property of many large-scale workloads is that they are dominated by
reads (e.g., they constitute 99.8\% of requests to the Facebook
graph~\cite{Venkataramani:2012:TFS:2213836.2213957}). Here, simple locks would
entail unnecessary overheads. Instead, the Reader-Writer (RW)
lock~\cite{Mellor-Crummey:1991:SRS} can be used to reduce the overhead among
processes that only perform reads in the critical section (CS). Initial RW
\emph{NUMA-aware} designs have recently been introduced~\cite{Calciu:2013:NRL},
but they do not address DM machines.
%
% Moreover, to retain flexibility, the lock should allow for high performance
% of not only readers but also writers.

In this work, we develop a lock that addresses the above challenges. Its core
concept is a modular design for adjusting performance to various types of
workloads. The lock consists of three key data structures. First, the distributed
counter (DC) indicates the number of readers or the presence of a writer in the
CS. Second, the distributed queue (DQ) synchronizes writers belonging to a given
element of the memory hierarchy (e.g., a rack). Finally, the distributed tree
(DT) binds together all queues at different levels of the memory hierarchy and
synchronizes writers with readers. Each of these three structures offers an
adjustable performance tradeoff, enabling high performance in various settings. 
DC can lower the latency of lock acquire/release performed by either readers or writers, DQ can be biased towards
improving either locality or fairness, and DT can increase the throughput of
either readers or writers.
The values of these parameters constitute a three dimensional space that is
illustrated in Figure~\ref{fig:space}. Each point is a specific lock design
with selected performance properties.

% Another challenge in today's large-scale processing is to develop an
% efficient lock for various analytics workloads dominated by reads (e.g.,
% 99.8\% of requests to the Facebook graph are read
% requests~\cite{Venkataramani:2012:TFS:2213836.2213957}). Here, simple locks
% would entail unnecessary synchronization overheads. Instead, the
% Reader-Writer (RW) lock~\cite{Mellor-Crummey:1991:SRS} is a more proper
% match.  It reduces the amount of synchronization for processes that only
% perform reads in the critical section (CS). Initial RW \emph{NUMA-aware} lock
% designs have recently been introduced~\cite{Calciu:2013:NRL}, but they do not
% address DM machines.

\goal{Advertise RMA and say we'll use it for our locks.}
\sloppy Most DM machines offer Remote Direct Memory Access
(RDMA)~\cite{recio2007remote}, a hardware scheme that removes the OS and the
CPU from the inter-node communication path. RDMA is the basis of many Remote
Memory Access (RMA)~\cite{fompi-paper} programming models. Among others, they offer a
Partitioned Global Address Space (PGAS) abstraction to the programmer and
enable low-overhead direct access to remote memories with put/get communication
primitives. RMA principles are used in various HPC languages and libraries:
Unified Parallel C (UPC)~\cite{upc}, Fortran 2008~\cite{fortran2008},
MPI-3~\cite{mpi3}, or SHMEM~\cite{shmem}.
We will illustrate how to utilize RMA in the proposed locks for DM
machines, addressing the above-mentioned challenges. In the following, we use
MPI-3 RMA but we keep our protocols generic and we discuss
(\cref{sec:discussion}) how other RMA languages and libraries can also be used.

% We will illustrate how to utilize RMA to design efficient locks for DM
% machines, addressing the above-mentioned challenges. In the following, we use
% MPI-3 RMA but we keep our protocols generic and we discuss
% (\cref{sec:discussion}) how other RMA languages and libraries can also be
% used.

\goal{State our contributions}
In summary, our key contributions are as follows:

\begin{itemize}
\item We develop a topology-aware distributed Reader-Writer lock
that enables various tradeoffs between fairness, throughput, latency,
and locality.
% %
% \item We develop a topology-aware distributed Reader-Writer lock
% that trades fairness for higher throughput.
% %
% \item We show that our schemes satisfy mutual exclusion,
% deadlock-freedom, and starvation-freedom.
%
% \item We improve the implementation of the utilized MPI-3 RMA library
% foMPI~\cite{fompi-paper} by seamlessly incorporating the distributed MCS lock
% into its design. The new lock implementation outperforms the previous scheme
% by 350\% on average. We also improve the implementation of an atomic
% operation Fetch-And-Replace provided by foMPI by utilizing hardware-supported
% Cray communication primitives.
%
\item We offer a topology-aware distributed MCS lock that accelerates the
state-of-the-art MPI-3 RMA codes~\cite{fompi-paper}.
\item We illustrate that our designs outperform the state-of-the-art in
throughput/latency (7.2x/6.8x on average) and that they accelerate
distributed hashtables used in key-value (KV) stores or graph processing.
\end{itemize}

\section{RMA AND LOCKS}
\label{sec:background}

\goal{Introduce the section}
We start by discussing RMA (\cref{sec:background_rma}), our tool to develop the
proposed locks. Next, we present traditional (\cref{sec:traditional_locks}) and
state-of-the-art (\cref{sec:state-of-the-art_locks}, \cref{sec:distributed_mcs})
locks that we use and extend.

% We start by presenting traditional (\cref{sec:traditional_locks}) and
% state-of-the-art (\cref{sec:state-of-the-art_locks}) locks that we use and
% extend in this work. Then, we discuss RMA (\cref{sec:background_rma}), our
% tool to develop the proposed locks.

% We start by presenting desired lock properties (\cref{sec:lock_properties})
% and then proceed to traditional (\cref{sec:traditional_locks}) and
% state-of-the-art (\cref{sec:state-of-the-art_locks}) locks that we use and
% extend in this work. Finally, we discuss RMA programming
% (\cref{sec:background_rma}).

\textbf{\textsf{Notation/Naming:}}
We denote the number of processes as $P$; we use the notion of a \emph{process}
as it occurs frequently in DM codes such as MPI~\cite{mpi3}.
Still, our schemes are independent of whether heavyweight processes or
lightweight threads are incorporated. Each process has a unique ID called the
\emph{rank} $\in \{1, ..., P\}$. A process in the CS is called \emph{active}. A
null pointer is denoted as $\emptyset$. Then, $N$ is the number of levels
of the memory hierarchy of the used machine. Here, the selection of the
considered levels depends on the user. For example, one can
only focus on the nodes connected with a network and racks that contain nodes
and thus $N=3$ (three levels: the nodes, the racks, and the whole machine). We
refer to a single considered machine part (e.g., a node) as an \emph{element}. 
We refer to a node that is a shared-memory cache-coherent domain connected to
other such domains with a non-coherent network as a \emph{compute node} (or just \emph{node}). One
compute node may contain smaller elements that are cache-coherent and together
offer \emph{non-uniform memory access (NUMA)}. We refer to such elements as
\emph{NUMA nodes}; an example NUMA node is a socket with a local DRAM.
%
% Finally, we denote processes that are acquiring or relasing the lock as
% \emph{acquirers} and \emph{releasers}.
%
We present symbols used in the paper in Table~\ref{tab:symbols}.

% Symbols used in the paper are presented in Table~\ref{tab:symbols}.

\begin{table}[h!]
%\vspace{-0.5em}
\centering
\footnotesize
%\scriptsize
\sf
\begin{tabular}{r|l}
\toprule
$P$ & Number of processes.\\
$p$ & Rank of a process that attempts to acquire/release a lock.\\
$N$ & Number of levels of the considered machine.\\
$N_i$ & Number of machine elements at level~$i$; $1 \leq i \leq N$.\\
$i$ & Index used to refer to the $i$th machine level.\\
$j$ & Index used to refer to the $j$th element at a given machine level.\\
\bottomrule
\end{tabular}
%\vspace{-0.5em}
\caption{Symbols used in the paper.}
\label{tab:symbols}
%\vspace{-2.0em}
%\vspace{-3.0em}
\end{table} 

\subsection{RMA Programming}
\label{sec:background_rma}

\goal{+ Describe RDMA \& RMA and show they're popular}

In RMA programming, processes communicate by directly accessing one another's
memories.  Usually, RMA is built over OS-bypass RDMA hardware for highest
performance.
%
% it offers a \emph{partitioned global address space (PGAS)} abstraction. 
%
RMA non-blocking \emph{put}s (writes to remote memories) and
\emph{get}s (reads from remote memories) offer low latencies, potentially outperforming
message passing~\cite{fompi-paper}.  Remote \emph{atomics} such as
compare-and-swap~\cite{mpi3,Herlihy:2008:AMP:1734069} are also available.
Finally, RMA \emph{flushes} ensure the consistency of data by synchronizing
respective memories. RDMA is provided in virtually all modern networks (e.g.,
IBM PERCS~\cite{arimilli2010percs}, IBM's on-chip Cell,
InfiniBand~\cite{IBAspec}, iWARP~\cite{iwarp}, and RoCE~\cite{roce}).
Moreover, numerous libraries and languages offer RMA features. Examples
include MPI-3 RMA~\cite{mpi3}, UPC~\cite{upc},
Titanium~\cite{hilfinger2005titanium}, Fortran 2008~\cite{fortran2008},
X10~\cite{x10}, or Chapel~\cite{chapel}. The number of RMA codes is
growing steadily, and RMA itself is being continually enhanced~\cite{besta2014fault, besta2015active}.

% We use MPI RMA for concreteness. Still, our
% schemes are generic and can be developed with any other RMA/PGAS library;
% see~\cref{sec:discussion} for a discussion. 

\textbf{\textsf{RMA Windows:}}
In RMA, each process explicitly exposes an area of its local memory as shared.
In MPI, this region is called a \emph{window}. Once shared, a window can be
accessed with puts/gets/atomics and synchronized with flushes.
We will refer to such an exposed memory in any RMA library/language as a window.

% In RMA, each process explicitly exposes an area of its local memory as
% shared.  In MPI RMA, this region is called a \emph{window}. Each process has
% a private and a public window. The process itself can modify both windows but
% all other remote processes can only make changes with puts/gets to the public
% one.  Once shared, a window can be accessed with various language-specific
% operations.

% Another important value, which is needed for RMA, is the rank of an
% MPI-process. It is used to identify each process in an MPI communication
% environment and therefore is unique for each process.

% Memory can be shared in different ways (e.g., MPI windows, UPC shared arrays,
% or Co-Arrays in Fortran 2008); details are outside the scope of this work.
% Once shared, memory can be accessed with various language-specific
% operations.

% \goal{+ Introduce the naming for origins and targets} % There are two basic
% types of RMA operations: \emph{communication} actions (often called
% \emph{accesses}; they transfer data between processes), and
% \emph{synchronization} actions (synchronize processes and guarantee memory
% consistency).  
%
% Here, we use \emph{origin} or \emph{target} to refer to a process that issues
% or is targeted by an RMA access. We use \emph{sender} and \emph{receiver} to
% name processes that exchange messages.

\textbf{\textsf{RMA Functions:}}
We describe the syntax/semantics of the used RMA calls in
Listing~\ref{lst:rma_calls}. All \texttt{int}s are 64-bit. For clarity, we
also use the \texttt{bool} type and assume it to be an \texttt{int}
that can take the 0 (\texttt{false}) or 1 (\texttt{true}) values, respectively.
Values
returned by \texttt{Get}/\texttt{FAO}/\texttt{CAS} are only valid after the
subsequent \texttt{Flush}. The syntax is simplified for clarity: we omit a
pointer to the accessed window (we use a single window).
We use an \emph{origin}/a
\emph{target} to refer to a process that issues or is targeted by an RMA call.

% We now shortly explain the utilized RMA functions. We show here simplified versions 
% of the functions because we omit the details for clarity.
% As stated before, there are two categories of RMA operations. First, we describe the 
% \emph{communication} actions: 
% 
% \begin{itemize}
% \item MPI\_Get fetches a piece of data on a remote MPI-process to the specified target location.
% This is only possible if the data is in the memory window of the target process. 
% 
% \item The counterpart of MPI\_Get is MPI\_Put which is essentially a remote write. Therefore, 
% it puts the given data to the specified location in the memory window of the target process.
% 
% \item The function MPI\_Accumulate executes a specified operation at 
% the target location with a given operand. We utilize the operations 
% MPI\_REPLACE that replaces the target location with the operand and 
% MPI\_SUM that adds the operand to the target location. 
% 
% \item An extension of MPI\_Accumulate is the function MPI\_Fetch\_and\_op that atomically 
% fetches first a piece of data at the target and then accumulates the operand to the value at the 
% target location with the declared operation. 
% 
% \item MPI\_Compare\_and\_swap first checks whether the target memory location is 
% equal to a compare value and if yes, swaps it with a given data. Moreover, it returns 
% the value of the target location before the swap.
% \end{itemize}
% 
% Below are the definitions of the functions: 

\begin{lstlisting}[float=h,label=lst:rma_calls,caption=The syntax/semantics of
the utilized RMA calls.]
/* |\underline{Common parameters:}| $target$: target's rank; $offset$: an offset
 * into $target$'s window that determines the location of the
 * targeted data; $op$: an operation applied to a remote piece of
 * data (either an atomic replace (REPLACE) or a sum (SUM)); 
 * $oprd$: the operand of an atomic operation $op$.*/

/* Place atomically $src\_data$ in $target$'s window.*/
void Put(int src_data, int target, int offset);

/* Fetch and return atomically data from $target$'s window.*/
int Get(int target, int offset);

/* Apply atomically $op$ using $oprd$ to data at $target$.*/
void Accumulate(int oprd, int target, int offset, MPI_Op op);

/* Atomically apply $op$ using $oprd$ to data at $target$
 * and return the previous value of the modified data.*/
int FAO(int oprd, int target, int offset, MPI_Op op);

/* Atomically compare $cmp\_data$ with data at $target$ and, if
 * equal, replace it with $src\_data$; return the previous data.*/
int CAS(int src_data, int cmp_data, int target, int offset);

/* Complete all pending RMA calls started by the calling process
 * and targeted at $target$.*/
void Flush(int target);
\end{lstlisting}

\subsection{Traditional Hardware-Oblivious Locks}
\label{sec:traditional_locks}

% We first present hardware-oblivious locks used in this work.
%

We now present hardware-oblivious locks used in this work.

\subsubsection{Reader-Writer (RW) Locks}
\label{sec:trad_rw_lock}

\goal{Describe RW locks}

Reader-Writer (RW) locks~\cite{Courtois:1971:CCL} distinguish between processes
that only perform reads when in the CS (\emph{readers}) and those that issue
writes (\emph{writers}). Here, multiple readers may simultaneously enter a
given CS, but only one writer can be granted access at a time, with no other
concurrent readers or writers. RW locks are used in OS kernels, databases, and
present in various HPC libraries such as MPI-3~\cite{mpi3}.

\subsubsection{MCS Locks}
\label{sec:trad_mcs_lock}

\goal{Describe MCS locks}

Unlike RW locks, the MCS lock (due to Mellor-Crummey and
Scott)~\cite{Mellor-Crummey:1991:ASS, Scott:2001:SQS:379539.379566, 292571}
does not distinguish between readers or writers. Instead, it only allows one
process $p$ at a time to enter the CS, regardless of the type of memory accesses
issued by $p$.
%
% We illustrate the MCS lock in Figure~\ref{fig:MCS}.
%
Here, processes waiting for the lock form a queue, with a process at the head
holding the lock.  The queue contains a single global pointer to its tail.
Moreover, each process in the queue maintains: (1) a local flag that signals if
it can enter the CS and (2) a pointer to its successor. To enter the queue, a
process $p$ updates both the global pointer to the tail and the pointer at its
predecessor so that they both point to $p$. 
A releasing process notifies its successor by changing the successor's local flag. 
The MCS lock reduces the amount of coherence traffic that limits the
performance of 
spinlocks~\cite{Anderson:1990:PSL:628891.628973}. Here, each process in the
queue spin waits on its local flag that is modified once by its predecessor.

\subsection{State-of-the-Art NUMA-Aware Locks}
\label{sec:state-of-the-art_locks}

We now discuss lock schemes that use the knowledge
of the NUMA structure of the underlying machine for more performance.
We will combine and 
extend them to DM domains, and enrich them with a family of adjustable parameters
for high performance with various workloads.

% \begin{figure}[h]
% 	\centering
% 	 \includegraphics[width=1\textwidth]{hMCS-eps-converted-to.pdf}
% 	 \caption{An example hMCS tree.}
% 	 \label{fig:hMCS}
% \end{figure}

\subsubsection{NUMA-Aware RW Locks}
\label{sec:numa_rw_lock}

\goal{Describe NUMA-aware RW-locks and explain why research is needed}

Many traditional RW locks (\cref{sec:trad_rw_lock}) entail performance penalties in
NUMA systems as they usually rely on a centralized structure that becomes
a bottleneck and entails high latency when accessed by processes from remote
NUMA elements.
Calciu et al.~\cite{Calciu:2013:NRL} tackle this issue with a flag on each NUMA
node that indicates if there is an active reader on that node.  This reduces
contention due to readers (each reader only marks a local flag) but may entail
additional overheads for writers that check for active readers. 

\subsubsection{Hierarchical MCS Locks}
\label{sec:hierarchical_locks}

\goal{Describe hierarchical locks and show why this is needed today}

Hierarchical locks tackle expensive lock passing described in~\cref{sec:intro}.
They trade fairness for higher throughput by ordering processes that enter the CS
to reduce the number of such passings.
Most of the proposed schemes address two-level NUMA
machines~\cite{Chabbi:2015:HPL:2688500.2688503, Dice:2011:FNL,
luchangco2006hclh, Radovic:2003:HBL}. Chabbi et al.~consider a multi-level NUMA
system~\cite{Chabbi:2015:HPL:2688500.2688503}. Here, each NUMA
hierarchy element (e.g., a socket) entails a separate MCS lock. To acquire the global
lock, a process acquires an MCS lock at each machine level.
This increases locality~\cite{tate2014programming} but reduces fairness: processes on the same NUMA
node acquire the lock consecutively even if processes on other nodes are waiting.

\subsection{Distributed RMA MCS Locks}
\label{sec:distributed_mcs}

\goal{Introduce the section.}
Finally, we present a distributed MCS (D-MCS) lock based on an MPI-3 MCS
lock~\cite{gropp2014using}. We will use it to accelerate state-of-the-art MPI
RMA library foMPI~\cite{fompi-paper} and as a building block of the proposed
distributed topology-aware RW and MCS locks (\cref{sec:rw_locks}).

\subsubsection{Summary and Key Data Structures}
\label{sec:mcs_data}

\goal{Tell how we use windows}

%\sloppy
Here, processes that wait for the D-MCS lock form a queue that may span multiple nodes. 
Each process maintains several globally visible variables. A naive approach
would use one window per variable. However, this would entail additional memory
overheads (one window requires $\Omega(P)$ storage in the worst
case~\cite{fompi-paper}).  Thus, we use one window with different offsets
determining different variables: a pointer to the next process in the MCS queue
(offset \texttt{NEXT}, initially $\emptyset$) and a flag indicating if a given
process has to spin wait (offset \texttt{WAIT}, initially \texttt{false}). 
%
% A selected process (rank \texttt{tail\_rank}) also maintains a rank of the
% process with the queue tail (offset \texttt{TAIL}, initially $\emptyset$). 
%
A selected process (rank \texttt{tail\_rank}) also maintains a pointer to a
process with the queue tail (offset \texttt{TAIL}, initially $\emptyset$). 
%
% Each pointer contains two fields: the rank of a targeted process and an
% address within this process' address space.  This reflects the hierarchical
% nature of distributed environment. 

% We now discuss the data structures that are used by the lock
% schemes.  They are all included in a RMA window and thus each process exposes
% their local instances to all other involved processes.
% %
% First, we use a pointer to the tail of the MCS queue, located on a selected
% node. The rank of this node is saved in the field \emph{rankTail}.
% % 
% Furthermore, we use another pointer at each process to the next process in
% queue. Finally, there is a variable (also at each process) that indicates
% whether a given process has to spin wait. We call it the \emph{blocked} value 
% of a process.
% 
% Each pointer consists of two variables. The first one is the process
% \emph{rank}: it identifies a specific process. The other one is an address
% within the given process' address space.
% %
% Such a hierarchical structure corresponds to the hierarchical nature of the
% distributed environment. 
% 
% Each lock must be initialized at the beginning of a program, only then can it
% be acquired and released. The allocated memory must be freed before the program
% termination. In Listing \ref{lst:mcs_constants} one can see the constants that 
% are used in the code. We now describe these four functions in more detail.

% \begin{lstlisting}[caption=Constants used for the MCS RMA lock,
% label=lst:mcs_constants]
% /* Offsets for the queueNode array */
% enum {NEXT = 0, WAIT = 1};
% \end{lstlisting}

\subsubsection{Lock Protocols}
%\subsection{RMA-MCS Lock Protocols}
\label{sec:mcs_implementation}

We now describe the protocols for acquire/release. We refer to respective
variables using their offsets in the window.
% The acquiring/releasing process has a rank $p$.

% \subsubsection{Lock Initialization}
% \label{sec:mcs_init}
% 
% \goal{description of initialization}
% 
% First, a process prepares its part of the memory window with the related
% variables so that  other processes can access it. This is not a part of the
% traditional MCS design and is required due to the utilization of RMA. After
% that, we initialize the above mentioned three variables that constitute the
% allocated memory region.  The value for indicating whether the process is
% blocked can be set to false. The initialization is only called once and
% therefore has no impact on the performance of acquiring and releasing the
% lock.

% \begin{lstlisting}[caption=(MCS-RMA, \cref{sec:mcs_init}) Key data structures
% and lock initialization.]
% mcs_rma_win[WAIT] = true;
% mcs_rma_win[NEXT] = $\null$;
% \end{lstlisting}

% \subsubsection{Lock Acquire (Listing~\ref{lst:mcs_acquire})}
% \label{sec:mcs_acquire}

\textbf{Lock Acquire (Listing~\ref{lst:mcs_acquire})}
First, $p$ atomically modifies \texttt{TAIL} with its own rank and fetches the
predecessor rank (Line~\ref{line:rma_mcs_fetch_pred}). If there is no
predecessor, it proceeds to the CS. Otherwise, it enqueues itself
(Line~\ref{line:rma_mcs_enqueue_itself}) and waits until its local
\texttt{WAIT} is set to \texttt{false}. \texttt{Flush}es ensure the data
consistency.

% After the initialization has been finished, each process can attempt to
% acquire the lock. The first step of the acquisition is to atomically fetch on
% the designated node the rank of the last in queue and replace it with its own
% rank. Fortunately, there is the powerful operation MPI\_Fetch\_And\_Op in
% MPI-3.0 that exactly does these two things in one go. If there is no
% predecessor, the lock is acquired successfully.  Otherwise, the processor has
% to put his rank into the \emph{next} field of his predecessor with the
% function MPI\_Put. Afterwards, it has to wait until its \emph{blocked} value
% is set to 1. Below is the code listed for acquiring a lock:

\begin{lstlisting}[float=h,label=lst:mcs_acquire,caption=Acquiring D-MCS.]
void acquire() {
  /* Prepare local fields. */
  Put($\emptyset$, $p$, NEXT);
  Put(true, $p$, STATUS);
  /* Enter the tail of the MCS queue and get the predecessor. */
  int pred = FAO($p$, tail_rank, TAIL, REPLACE);|\label{line:rma_mcs_fetch_pred}| 
  Flush(tail_rank); /* Ensure completion of FAO. */
  if(pred != $\emptyset$) { /* Check if there is a predecessor. */
    /* Make the predecessor see us. */
	  Put($p$, pred, NEXT); Flush(pred);|\label{line:rma_mcs_enqueue_itself}|
    bool waiting = true;
	  do { /* Spin locally until we get the lock. */
		  waiting = Get($p$, WAIT); Flush($p$);
	  } while(waiting == true); } }
\end{lstlisting}

% \subsubsection{Lock Release (Listing~\ref{lst:mcs_release})}
% \label{sec:mcs_release}

\textbf{Lock Release (Listing~\ref{lst:mcs_release})}
%
% We present this scheme in Listing~\ref{lst:mcs_release}. 
%
First, $p$ checks if it has a successor in the queue
(Line~\ref{line:rma_mcs_release_check_succ}).  If there is none, it atomically
verifies if it is still the queue tail
(Line~\ref{line:rma_mcs_release_check_tail}); if yes, it sets \texttt{TAIL} to
$\emptyset$. Otherwise, $p$ waits for a process that has
modified \texttt{TAIL} to update its \texttt{NEXT} field
(Lines~\ref{line:rma_mcs_release_wait_start}-\ref{line:rma_mcs_release_wait_end}).
If there is a successor, the lock is passed with a single \texttt{Put}
(Line~\ref{line:rma_mcs_release_notify}). 

\begin{lstlisting}[label=lst:mcs_release,float=h,caption=Releasing D-MCS.]
void release() {
  int succ = Get($p$, NEXT); Flush($p$);
  if(succ == $\emptyset$) { |\label{line:rma_mcs_release_check_succ}| 
    /* Check if we are waiting for the next proc to notify us.*/
    int curr_rank = CAS($\emptyset$, $p$, tail_rank, TAIL); |\label{line:rma_mcs_release_check_tail}|
    Flush(tail_rank);    
    if($p$ == curr_rank)
      return; /* We are the only process in the queue. */
    do { /* Wait for a successor. */ |\label{line:rma_mcs_release_wait_start}|
      successor = Get($p$, NEXT); Flush($p$);
    } while (successor == $\emptyset$); |\label{line:rma_mcs_release_wait_end}|
  }
  /* Notify the successor. */
  Put(0, successor, WAIT); Flush(successor);|\label{line:rma_mcs_release_notify}|}
\end{lstlisting}

\begin{figure*}
\centering
\includegraphics[width=1\textwidth]{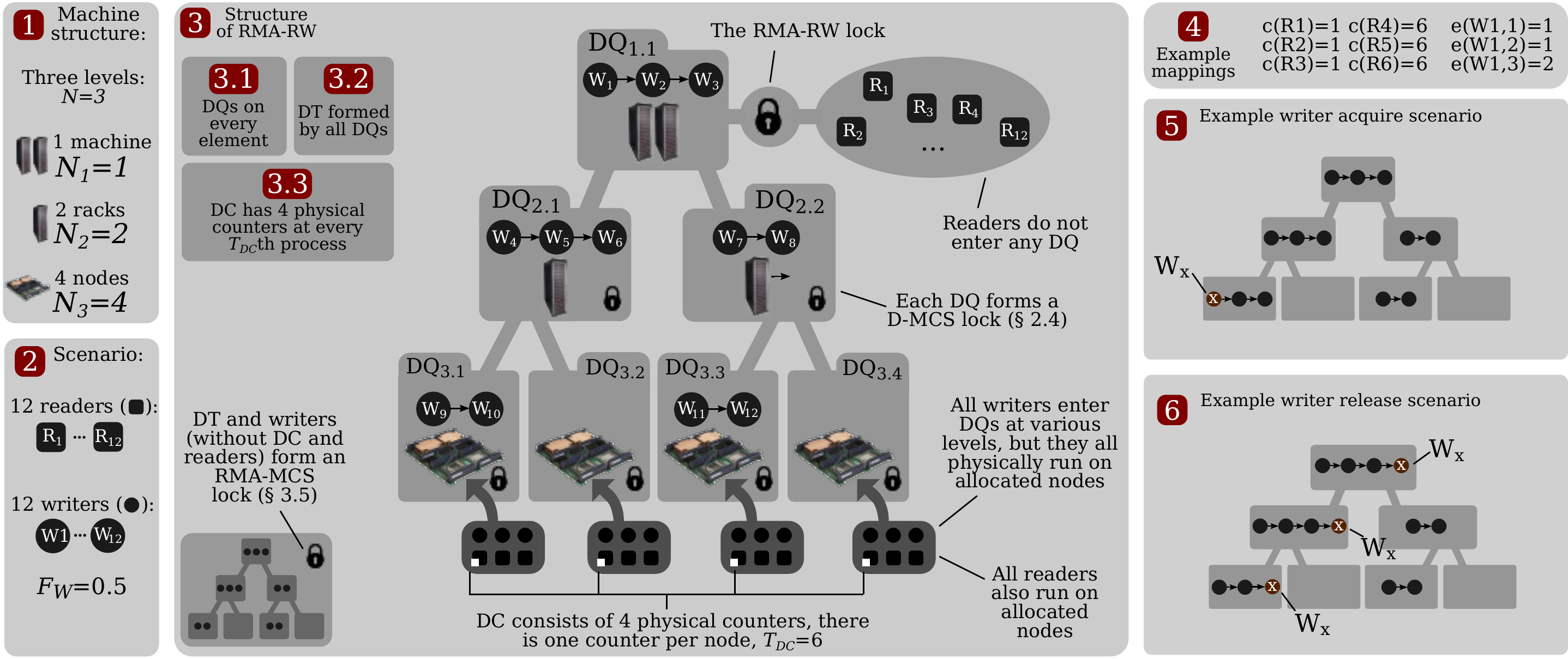}
\caption{An example RMA-RW on a three-level system.}
\label{fig:structures}
\end{figure*}

\section{DISTRIBUTED RMA RW LOCKS}
\label{sec:rw_locks}

We now present a distributed \emph{topology-aware} RW lock (RMA-RW) for
scalable synchronization and full utilization of parallelism in workloads
dominated by reads. 
We focus on the RW semantics as the key part of the introduced lock.
Symbols specific to RMA-RW are presented in Table~\ref{tab:symbols_rma-rw}.

% Next, we present a distributed \emph{topology-aware} RW lock (RMA-RW)
% for scalable synchronization and full utilization of  
% parallelism in workloads dominated by read accesses. 
% %
% We present symbols specific to RMA-RW in Table~\ref{tab:symbols_rma-rw}.

\macb{Lock Abbreviations}
We always refer to the proposed topology-aware distributed RW and MCS lock
as RMA-RW and RMA-MCS, respectively. Both RMA-RW and RMA-MCS use as
their building block a simple distributed topology-oblivious
MCS lock (\cref{sec:distributed_mcs}) denoted as D-MCS.

\macb{Example}
In the whole section, we will use the example shown in~Figure~\ref{fig:structures}.
Here, $N=3$ and the considered levels are: compute nodes, racks,
and the whole machine.

% Symbols used in the paper are presented in Table~\ref{tab:symbols}.

\begin{table}[h!]
%\vspace{-0.5em}
\centering
\footnotesize
%\ssmall
\sf
\begin{tabular}{r|l}
\toprule
$T_{DC}$ & The \emph{Distributed Counter} threshold (\cref{sec:counters}).\\
% $T_{DC}$ & Number of parts of a DC counter (\cref{sec:counters}).\\
$T_{L,i}$ & The \emph{Locality} threshold at level~$i$ (\cref{sec:qnode}).\\
% $T_{L,i}$ & The maximum number of lock acquires within one element at level~$i$ (\cref{sec:qnode}).\\
$T_{R}$ & The \emph{Reader} threshold (\cref{sec:hnode}).\\
$T_{W}$ & The \emph{Writer} threshold; $T_{W} = \prod_{i=1}^{N} T_{L,i}$ (\cref{sec:hnode}).\\
% $T_{RW}$ & The maximum number of consecutive lock acquires of readers/writers (\cref{sec:hnode}).\\
$c(p)$ & Mapping from a process $p$ to its physical counter (\cref{sec:counters}).\\
% $c(p)$ & Rank of a process with a physical counter associated with a process $p$.\\
% $e(p,i)$ & Mapping from a process $p$ to its home machine element at level~$i$; $1 \leq e(p,i) \leq N_i$.\\
$e(p,i)$ & Mapping from a process $p$ to its home machine element at level~$i$ (\cref{sec:qnode}).\\
$F_W$ & The fraction of writers in a given workload (the fraction of readers: $1-F_W$).\\
\bottomrule
\end{tabular}
%\vspace{-0.5em}
\caption{Symbols used in RMA-RW.} 
\label{tab:symbols_rma-rw}
%\vspace{-2.0em}
%\vspace{-3.0em}
\end{table}

% Next, we present the design of a \emph{distributed topology-aware
% Reader-Writer lock} (RMA-RW).  We start by explaining the intuition behind
% the scheme (\cref{sec:concept}). Later, we discuss the utilized data
% structures (\cref{sec:key_data}) and provide more details on how the lock is
% acquired and released (\cref{sec:rw_implementation}).

\subsection{Design Summary and Intuition}
\label{sec:concept}

% \impr{We now present how the above data structures play together in the
% acquire and release protocols.  A writer always starts at the leaf of the
% tree (level $N$) both for acquiring and releasing.  At any level $i$ where $2
% \le i \le N$, it proceeds up the tree executing a protocol for a partial
% acquire/release of the respective part of the tree
% (\cref{sec:writer_acquire_n}, \cref{sec:writer_release_n}).  At this level,
% it competes with other writers that attempt to enter the CS.  For level $1$
% If it reaches level 1, it executes a different protocol for locking or
% releasing the whole lock (\cref{sec:writer_acquire_1},
% \cref{sec:writer_release_1}). Here, it also competes with readers. Readers do
% not follow such a hierarchy and thus have a single acquire
% (\cref{sec:reader_acquire}) and release (\cref{sec:reader_release})
% protocol.}

As explained in~\cref{sec:intro}, RMA-RW consists of three types of core data
structures: distributed queues (DQs), a distributed tree (DT), and a
distributed counter (DC). They are illustrated in Figure~\ref{fig:structures}.
First, every machine element (at each considered level) has an associated DQ
and thus a D-MCS lock \emph{local} to this element (as opposed to the
\emph{global} RMA-RW lock). 
In our example, every node, rack, and the whole machine have their own DQ (and
thus a local MCS lock).  Note that some DQs that are associated with elements
such as nodes are not necessarily distributed, but we use the same name for
clarity.
%
% For example, if $N=3$ and the considered elements are compute nodes, racks,
% and the whole machine, then there is one DQ (and thus an MCS lock) running on
% every compute node, every rack, and the whole machine (note that some DQs
% that are associated with elements such as compute nodes are not necessarily
% distributed, but we use the same name for clarity).
%
Second, all the DQs form a DT that corresponds to the underlying memory
hierarchy, with one DQ related to one tree vertex. For example, DQs associated
with nodes that belong to a given rack $r$ constitute vertices that are children of
a vertex associated with a DQ running on rack $r$. 
Third, DC counts active readers and writers and consists of several physical
counters located on selected processes. 
DT on its own (without DC and any readers) constitutes RMA-MCS.

% Writers that execute on a given machine element (e.g., a rack) belong to one
% DQ assigned to this element. All such DQs are combined in a single DT that
% spans the whole machine.
% 
% In RMA-RW, each writer  that executes on a given machine element (e.g., a
% rack) belongs to one DQ assigned to this element. All such DQs are combined
% in a single DT that spans the whole machine. Contrarily, readers do not enter
% any queue for lightweight synchronization.

% use topology-aware MCS-based queues that maximize locality while readers do
% not enter any queue for lightweight synchronization. A writer that
% acquires/releases RMA-RW must acquire/release up to $N$ MCS locks (one per
% memory hierarchy level), similarly to the strategy by Chabbi et al.
% (\cref{sec:hierarchical_locks}). Contrarily, readers only synchronize with
% writers that acquired the last MCS lock. Finally, RMA-RW enables adjusting
% the performance of readers and writers by trading the throughput and latency
% of either one in exchange for the other, depending on which one is more
% important in a particular workload. 

% We illustrate an example in Figure~\ref{fig:intuition}. Every considered
% memory hierarchy element (e.g., a socket or a node) has an associated
% distributed MCS lock and thus a queue; all such queues form a tree of locks.
% At a given tree level, lock passing can be biased towards trading fairness
% for higher throughput. 
% 
% The number of readers and the presence of writers in the CS is indicated by
% the same system-wide distributed counter.

\macb{Writers }
A writer that wants to acquire a lock starts at a leaf of DT located at the
lowest level~$N$ (a node in our example). At any level~$i$ ($2 \le i \le N$),
it acquires a local D-MCS lock that corresponds to a subtree of D-MCS locks (and
thus DQs) rooted at the given element. Here, it may compete with other writers.
When it reaches level~1, it executes a different protocol for acquiring the
whole RMA-RW lock. Here, it may also compete with readers. RMA-RW's
locality-aware design enables a \emph{shortcut}: some writers stop before
reaching level~1 and directly proceed to the CS. This happens if a lock
is passed within a given machine element. 

\macb{Readers }
Readers do not enter DQs and DT and thus have a single acquire protocol. This
design reduces synchronization overhead among readers.

\subsection{Key Data Structures}
\label{sec:key_data}

We now present the key structures in more detail. 
%
% the pseudocode can be found in Listing~\ref{lst:structures}.  
%
% Similarly to RMA-MCS, they occupy one RMA window to reduce storage overheads.

% The design of the RMA-RW lock is an interplay between three key data
% structures, each of them serving a different purpose. The \emph{distributed
% counter (DC)} indicates the number of readers or the presence of a writer in a
% CS. The \emph{distributed queue (DQ)} synchronizes writers at a given level of
% the memory hierarchy. Finally, the \emph{distributed tree (DT)} binds together
% queues at different levels of the memory hierarchy. We illustrate these data
% structures in Figure~\ref{fig:key-rw-structures}. This modular design
% facilitates the analysis of RMA-RW.

% We use three key data structures in our implementation. One of them is needed
% to control the struggle between readers and writers. Ideally, it should be
% decentralized since we want that not all readers have to check the same
% memory location. Moreover, it should make the readers topology-aware.
% Another structure is in charge of keeping the fairness between the writers
% because they only can execute the critical section one after another. Lastly,
% we want that the writers are NUMA-aware, too. For that purpose we have the
% last data structure. It uses the data structure that keeps the fairness
% between the writers multiple times -- one per NUMA domain. This should reduce
% the lock passing latency between writers. We now describe each one of them in
% more detail.

\subsubsection{Distributed Counter (DC)}
\label{sec:counters}

\goal{explain how the modes READ and WRITE work}
DC maintains the number of active readers or writers. It enables an adjustable
performance tradeoff that accelerates readers or writers. For this, one DC
consists of multiple physical counters, each maintained by every $T_{DC}$th
process; $T_{DC}$ is a parameter selected by the user. To enter the CS, a
reader $p$ increments only one associated physical counter while a writer must
check each one of them. Thus, selecting more physical counters (smaller
$T_{DC}$) entails lower reader latency (as each reader can access a counter
located on a closer machine element) and contention (as each counter is
accessed by fewer readers). Yet, higher $T_{DC}$ entails lower latency for a
writer that accesses fewer physical counters.

A physical counter associated with a reader $p$ is located at a rank $c(p)$;
$c(\cdot) \in \{1, ..., P\}$ can be determined at compile- or run-time. In a
simple hardware-oblivious scheme, one can fix $c(p) = \left\lceil p / T_{DC}
\right\rceil$.  For more performance, the user can locate physical counters in
a topology-aware way.  For example, if the user allocates $x$ processes/node
and a node $s$ hosts processes with $x$ successive ranks starting from
$(s-1)x+1$, then setting $T_{DC} = kx$ in the above formula results in storing
one physical counter every $k$th node.
%
%ranks $\{ (s-1)x + 1, (s-1)x + 2, ..., sx \}$, then
%
This can be generalized to any other machine element. 

To increase performance, we implement each physical counter as two 64-bit
fields that count the readers (assigned to this counter) that arrived and
departed from the CS, respectively. This facilitates obtaining the number of
readers that acquired the lock since the last writer and reduces contention
between processes that acquire and release the lock. We dedicate one bit of the
field that counts arriving readers to indicate whether the CS of RMA-RW is in
the \texttt{READ} mode (it contains readers) or the \texttt{WRITE} mode (it
contains a writer).

\textbf{\textsf{RMA Design of DC:}}
Each physical counter occupies two words with offsets \texttt{ARRIVE} (for
counting arriving readers) and \texttt{DEPART} (for counting departing
readers); physical counters together constitute an RMA window. 

\subsubsection{Distributed Queue (DQ)}
\label{sec:qnode}

DQ orders writers from a single element of the machine that attempt
to enter the CS. DQs from level~$i$ have an associated threshold $T_{L,i}$
that determines the maximum number of lock passings between writers running on
a machine element from this level before the lock is passed to a process from a
different element. We use a separate threshold $T_{L,i}$ for each~$i$ because
some levels (e.g., racks) may need more locality (a higher threshold) than
others (e.g., nodes) due to expensive data transfers. This design enables an
adjustable tradeoff between fairness and throughput at each level. 

DQ extends D-MCS in that the local flag that originally signals whether a
process can enter the CS now becomes an integer that carries (in the same RMA
operation) the number of past lock acquires within a given machine element. We
use this value to decide whether to pass the lock to a different
element at a given level~$i$ (if the value reaches $T_{L,i}$) or not
(if the value is below $T_{L,i}$).

% The design of the DQ is based on the MCS queue but we expand the use of the
% field \emph{status} so that each process can pass the count of locks that
% have been acquired in this queue to his successor.  Therefore, we introduce a
% value \emph{WAIT} that is the maximum value of a 64-bit integer.  Everything
% smaller than that number indicates that the process does not have to wait for
% the lock. We use this extension to introduce a threshold for each DQ. Every
% time the count gets bigger than the threshold, the lock of the queue one
% level above in the hierarchy is released as well. 

\textbf{\textsf{RMA Design of DQ:}}
All DQs at a given level constitute an RMA window. Respective offsets in the window
are as follows: \texttt{NEXT} (a rank of the next process in the queue),
\texttt{STATUS} (an integer that both signals whether to spin wait and carries
the number of past lock acquires in the associated machine element), and
\texttt{TAIL} (a rank of the process that constitutes the current tail of the
queue). \texttt{TAIL} in DQ at level~$i$ associated with $j$th element is stored on
a process \texttt{tail\_rank[$i$,$j$]}.

\subsubsection{Distributed Tree of Queues (DT)}
\label{sec:hnode}

DT combines DQs at different memory hierarchy levels into a single structure.
This enables $p$ to make progress in acquiring/releasing RMA-RW by moving
from level~$N$ to level~1. Then, at the tree root, writers synchronize with
readers. Specifically, the lock is passed from writers to readers (if there are
some waiting) when the total number of lock passings between writers reaches a
threshold $T_W$. In our design, $T_W = \prod_{i=1}^{N} T_{L,i}$.  To avoid
starvation of writers, we also introduce a threshold $T_R$ that is the maximum
number of readers that can enter the CS consecutively before the lock is passed
to a writer (if there is one waiting). Increasing $T_R$ or $T_{W}$ improves the
throughput of readers or writers because more processes of a given type can
enter the CS consecutively.

While climbing up DT, a writer must determine the next DQ (and thus D-MCS) to
enter.  This information is encoded in a mapping $e(\cdot,\cdot)$ and structure
\texttt{tail\_rank[$i$,$j$]}.  $e(p,i)$ $\in \{ 1, ..., N_i \}$ returns the ID
of a machine element associated with a process $p$ at level~$i$.  An expression
\texttt{tail\_rank[$i$,$e(p,i)$]} returns the rank of a process that points to
the tail of a DQ at level~$i$ within a machine element assigned to $p$. This
enables $p$ to enter D-MCS at the next level on the way to the CS.  Similarly
to $c(p)$, $e(p,i)$ can be determined statically or dynamically.

% $e(p,i)$: $p$ uses \texttt{tail\_rank[$i$,$j$]} and $e(p,i)$ to locate the
% tail of a parent DQ. Thus, it does not come with any RMA structures.

% Similarly to $c(p)$, $e(p,i)$ can be determined statically or dynamically.
% We will often use an expression \texttt{tail\_rank[$i$,$e(p,i)$]} that is the
% rank of a process that points to the tail of a DQ at level~$i$ within a
% machine element where $p$ is executing.

% The level~1 threshold ($T_{L,1}$) is special as it is associated with the
% whole machine. When it is reached, the lock is passed to the waiting readers.
% Note that it is only reached when $T_{L,i}$ for all other levels are reached.
% Thus, the total number of lock passings between writers before the lock goes
% to readers is $\prod_{i=1}^{N} T_{L,i}$. This prevents reader starvation.

Depending on $T_{L,i}$, some writers do not have to climb all DT levels and can
proceed directly to the CS.  Thus, we further extend the \texttt{STATUS} field
used in DQ with one more special value \texttt{ACQUIRE\_PARENT}.  This
indicates that $p$ cannot directly enter the CS and should continue up DT.

\subsubsection{Discussion on the Status Field}

A central part of DQ and DT is the \texttt{STATUS} field that enables processes
to exchange various additional types of information in a single RMA
communication action, including: (1) if a lock mode changed (e.g., from
\texttt{READ} to \texttt{WRITE}), (2) if a given process should acquire a lock
at a higher DT level, (3) if a given process can enter the CS, and (4) the
number of past consecutive lock acquires.
Two selected integer values are dedicated to indicate (1) and (2). All the
remaining possible values indicate that the given process can enter the CS (3);
at the same time the value communicates (4).

\subsection{Distributed Reader-Writer Protocol}

We now illustrate how the above data structures play together in the acquire
and release protocols. A writer starts at the leaf of DT 
(level~$N$) both for acquiring and releasing. At any level~$i$ ($2 \le i \le
N$), it proceeds up the tree executing a protocol for a partial acquire/release
of the respective part of the tree (\cref{sec:writer_acquire_n},
\cref{sec:writer_release_n}). At level~1, it executes a different
protocol for locking or releasing the whole lock (\cref{sec:writer_acquire_1},
\cref{sec:writer_release_1}). Readers do not follow such a hierarchy and thus
have single acquire (\cref{sec:reader_acquire}) and release
(\cref{sec:reader_release}) protocols.

\subsubsection{Writer Lock Acquire: Level $N$ to $2$ (Listing~\ref{lst:writer_acquire_n})}
\label{sec:writer_acquire_n}

\goal{description of writer acquisition}

% We present the protocol in Listing~\ref{lst:writer_acquire_n}. 

\textbf{\textsf{Intuition:}}
$p$ enters the DQ associated with a given level~$i$
and its home element $e(p,i)$; it then waits for the update from its
predecessor. If the predecessor does not have to hand over the lock to a process from
another element (i.e., has not reached the threshold $T_{L,i}$), the lock is
passed to $p$ that immediately enters the CS. Otherwise, $p$ moves to level~$i-1$.

\noindent
\textbf{\textsf{Details:}}
$p$ first modifies its \texttt{NEXT} and \texttt{STATUS} to reflect it
spin waits at the DQ tail
(Lines~\ref{line:writer_acq_n_local1}-\ref{line:writer_acq_n_local2}).  Then,
it enqueues itself (Line~\ref{line:writer_far_n}).  If there is a predecessor
at this level, $p$ makes itself visible to it with a \texttt{Put}
(Line~\ref{line:writer_put_n}) and then waits until it obtains the lock. While
waiting, $p$ uses \texttt{Get}s and \texttt{Flush}es to check for any updates
from the predecessor. If the predecessor reached $T_{L,i}$ and released the
lock to the parent level, $p$ must itself acquire the lock from level~$i-1$
(Line~\ref{line:writer_acq_n_up}). Otherwise, it can directly enter the CS
as the lock is simply passed to it
(Line~\ref{line:writer_acq_n_go}). If there is no predecessor at level~$i$, $p$
also proceeds to acquire the lock for level~$i-1$
(Line~\ref{line:writer_acq_n_up}).

% In Listing 4 the pseudo-code for this function is given. On each arbitrary
% level i between 2 and N, a process enqueues himself by using the function
% MPI_Fetch_and_replace (line 6). After that we need to call the function
% MPI_Win_flush (line 8) in order to make sure that the fetch and op has fin-
% ished. The first in the queue on level i holds the local MCS lock and
% proceeds to acquire the lock for the level i - 1.  If there is a predecessor
% on level i, the process makes itself visible to the predecessor by using the
% function MPI_Put (line 14) and then waits until the lock gets passed. While
% waiting the process needs to synchronize its private copy of the window with
% the public so that it will notify any changes to it. After that, there are
% two possibilities on how to pro- ceed: if the predecessor has hit the
% threshold for level i and released the lock to the parent level, the waiting
% process has to acquire it again or the acquisition is finished and the crit-
% ical section can be entered.  If there is no predecessor on level i, the
% process proceeds to acquire the lock on level i - 1.

\begin{lstlisting}[float=h,caption=Acquiring the RMA-RW lock by a writer; levels $N$
to $2$.,label=lst:writer_acquire_n]
void writer-acquire<$i$>() {
  Put($\emptyset$, $p$, NEXT);|\label{line:writer_acq_n_local1}|
  Put(WAIT, $p$, STATUS); Flush($p$);|\label{line:writer_acq_n_local2}|
  /* Enter the DQ at level $i$ and in this machine element. */
  int pred = FAO($p$, tail_rank[$i$,$e(p,i)$], TAIL, REPLACE);|\label{line:writer_far_n}|
  Flush(tail_rank[$i$,$e(p,i)$]);|\label{line:writer_flush_n}|
  if(pred != $\emptyset$) { 
    Put($p$, pred, NEXT); Flush(pred); /* pred sees us. */ |\label{line:writer_put_n}|
    int status = WAIT;
    do { /* Wait until pred passes the lock. */
      status = Get($p$, STATUS); Flush($p$);
    } while(status == WAIT); 
    /* Check if pred released the lock to the parent level. This
       would happen if |$T_{L,i}$| is reached. */
    if(status != ACQUIRE_PARENT) { 
      /* |$T_{L,i}$| is not reached. Thus, the lock is passed to
         $p$ that directly proceeds to the CS. */
      return; /* The global lock is acquired. */|\label{line:writer_acq_n_go}|
    }
  }
  /* Start to acquire the next level of the tree.*/
  Put(ACQUIRE_START, $p$, STATUS); Flush($p$);
  writer-acquire<$i-1$>();|\label{line:writer_acq_n_up}|}
\end{lstlisting}

\subsubsection{Writer Lock Release: Level $N$ to $2$ (Listing~\ref{lst:writer_release_n})}
\label{sec:writer_release_n}

\textbf{\textsf{Intuition:}}
$p$ passes the lock within $e(p,i)$ if
there is a successor and $T_{L,i}$ is not yet reached. Otherwise, it releases
the lock to the parent level~$i-1$, leaves the DQ, and informs any new
successor that it must acquire the lock at level~$i-1$.

\noindent
\textbf{\textsf{Details:}}
$p$ first finds out whether it has a successor. If there is one and $T_{L,i}$
is not yet reached, the lock is passed to it with a \texttt{Put}
(Line~\ref{line:release_put_n}). If $T_{L,i}$ is reached, $p$ releases the lock for
this level and informs its successor (if any) that it has to acquire the lock
at level~$i-1$. If there is no known successor, it checks atomically if some
process has already entered the DQ at level~$i$
(Line~\ref{line:release_cas_n}). If so, the releaser waits for the successor to
make himself visible before it is notified to acquire the lock at
level~$i-1$.

% Pseudo-code for this function is provided in Listing
% \ref{lst:writer_release_n}.  For the release of a writer lock on an arbitrary
% level i between 2 and N, a process first has to find out whether there is a
% successor. If there is one and the threshold for level i is not already
% reached, the lock can be passed to him by executing an MPI\_Put (line
% \ref{line:release_put_n}). 
% 
% In case that there is no successor or the threshold is reached, the writer
% releases the lock for this level and tries to inform his successor that it
% has to acquire the level i - 1 lock. 
% 
% If there is no known successor, we need to make sure that no one already
% enqueued himself in the queue on this level. Therefore, the process checks
% whether it is still the only one in the queue by using the
% MPI\_Compare\_and\_swap (line \ref{line:release_cas_n}) operation on the
% tail. If it indeed is the only one, the release is done. Otherwise, the
% process has to wait until the successor makes himself known. This is only for
% a short moment since a process only executes an MPI\_Win\_flush between
% enqueueing and making itself visible to its successor. Then the successor of
% the process gets the notice that it needs to acquire the parent level lock. 

\begin{lstlisting}[float=h,caption=Releasing an RMA-RW lock by a writer; levels
$N$ to $2$., label=lst:writer_release_n]
void writer-release<$i$>() {
  /* Check if there is a successor and get the local status. */
  int succ = Get($p$, NEXT); 
  int status = Get($p$, STATUS); Flush($p$);
  if(succ != $\emptyset$ && status < $T_{L,i}$) {
    /* Pass the lock to succ at level i as well as the number
       of past lock passings within this machine element. */
    Put(status + 1, succ, STATUS); Flush(succ); return;|\label{line:release_put_n}|
  } 
  /* There is no known successor or the threshold at level $i$ is 
     reached. Thus, release the lock to the parent level. */
  writer-release<$i-1$>();
  if(succ == $\emptyset$) {
    /* Check if some process has just enqueued itself. */
    int curr_rank = CAS($\emptyset$, $p$, tail_rank[$i$,$e(p,i)$], TAIL); |\label{line:release_cas_n}|
    Flush(tail_rank[$i$,$e(p,i)$]);
    if($p$ == curr_rank) { return; } 
    do { /* Otherwise, wait until succ makes itself visible. */
      succ = Get($p$, NEXT); Flush($p$);
    } while(succ == $\emptyset$);
  }
  /* Notify succ to acquire the lock at level $i-1$. */
  Put(ACQUIRE_PARENT, succ, STATUS); Flush(succ); }
\end{lstlisting}

 % int threshold = Get(tail_rank[$i$,$e(p,i)$], THRESHOLD);

\subsubsection{Writer Lock Acquire: Level 1 (Listing~\ref{lst:writer_acquire_1})}
\label{sec:writer_acquire_1}

% We present the scheme in Listing~\ref{lst:writer_acquire_1}. 

\textbf{\textsf{Intuition:}}
This scheme is similar to acquiring the lock at lower levels
(\cref{sec:writer_acquire_n}). However, the predecessor may
notify $p$ of the \emph{lock mode change} that enabled readers to
enter the CS, forcing $p$ to acquire the lock from the readers.

\noindent
\textbf{\textsf{Details:}}
$p$ first tries to obtain the lock from a predecessor
(Lines~\ref{line:writer_acq_1_get_from_pred_start}-\ref{line:writer_acq_1_get_from_pred_end}).
If there is one, $p$ waits until the lock is passed. Still, it can happen that
the predecessor hands the lock over to the readers
(Line~\ref{line:writer_acq_1_mode_change}). Here, $p$ 
changes the mode back to \texttt{WRITE} before
entering the CS (Line~\ref{line:acquire_cacc1_1}); this function checks each
counter to verify if there are active readers. If not, it switches
the value of each counter to \texttt{WRITE} (see
Listing~\ref{lst:manipulate_counters}).
If there is no predecessor (Line~\ref{line:writer_acq_1_no_pred}),
$p$ tries to acquire the lock from the readers by changing the mode to
\texttt{WRITE} (Line~\ref{line:acquire_cacc2_1}). 

% After changing all counters, $p$ waits for any remaining readers to leave the
% CS. A check for active readers is performed twice to detect a potential
% reader that might acquire the lock from a counter that is not yet swapped to
% \texttt{WRITE}. 

% \begin{lstlisting}[float=h,caption=Function that checks each counter for active
% readers., label=lst:check_counter]
% void check_counters() {
%   int i;
%   int64_t arr_cnt, dep_cnt;
% 
%   for(i = 0; i < world_size; i += COUNTER_DISTANCE) 
%   {
%     do 
%     {
%       /* We get the two counts of this counter */
%       arr_cnt = Get(i, ARRIVE);
%       dep_cnt = Get(i, DEPART);
%       Flush(i);
%       /* We check whether no reader is active when we already changed the counters to WRITE */
%       if(arr_cnt - dep_cnt - WRITE == 0)
%       {
%         break;
%       }
%     } while(arr_cnt - dep_cnt > 0);
%     /* There are no active readers for this counter */
%   }
% }
% \end{lstlisting}

\begin{lstlisting}[float=h,caption=Functions that manipulate counters., label=lst:manipulate_counters]
/****** Change all physical counters to the WRITE mode ******/
void set_counters_to_WRITE() { 
  /* To simplify, we use one counter every $T_{DC}$th process.*/
  for(int $p$ = 0; $p$ < $P$; $p$ += $T_{DC}$) {
    /* Increase the arrival counter to block the readers.*/
    Accumulate(INT64_MAX/2, $p$, ARRIVE, SUM); Flush($p$);
} }

/***************** Reset one physical counter *****************/
void reset_counter(int rank) {
  /* Get the current values of the counters.*/
  int arr_cnt = Get(rank, ARRIVE), dep_cnt = Get(rank, DEPART); 
  Flush(rank);
  /* Prepare the values to be subtracted from the counters.*/
  int sub_arr_cnt = -dep_cnt, sub_dep_cnt = -dep_cnt;

  /* Make sure that the WRITE is reset if it was set.*/
  if(arr_cnt >= INT64_MAX/2) {
    sub_arr_cnt -= INT64_MAX/2;
  }
  /* Subtract the values from the current counters.*/
  Accumulate(sub_arr_cnt, rank, ARRIVE, SUM);
  Accumulate(sub_dep_cnt, rank, DEPART, SUM); Flush(rank);
}

/***************** Reset all physical counters ****************/
void reset_counters() { 
  for(int $p$ = 0; $p$ < $P$; $p$ += $T_{DC}$) { reset_counter($p$); } }
\end{lstlisting}

\begin{lstlisting}[float=h,caption=Acquiring an RMA-RW lock by a writer; level
1., label=lst:writer_acquire_1]
void writer-acquire<1>() {
  Put($\emptyset$, $p$, NEXT); Put(WAIT, $p$, STATUS);|\label{line:writer_acq_1_get_from_pred_start}|
  Flush($p$); /* Prepare to enter the DQ.*/
  /* Enqueue oneself to the end of the DQ at level 1.*/
  int pred = FAO($p$, tail_rank[1,$e(p,1)$], TAIL, REPLACE);
  Flush(tail_rank[1,$e(p,1)$]);
  
  if(pred != $\emptyset$) { /* If there is a predecessor...*/
    Put($p$, pred, NEXT); Flush(pred);
    int curr_stat = WAIT;
    do { /* Wait until pred notifies us.*/
      curr_stat = Get($p$, STATUS); Flush($p$);
    } while (curr_stat == WAIT); 
    if(curr_stat == MODE_CHANGE) { /* The lock mode changed...*/|\label{line:writer_acq_1_mode_change}|
      /* The readers have the lock now; try to get it back.*/
      set_counters_to_WRITE(); |\label{line:acquire_cacc1_1}|
      Put(ACQUIRE_START, $p$, STATUS); Flush($p$);
    } } |\label{line:writer_acq_1_get_from_pred_end}|
  else { /* If there is no predecessor...*/|\label{line:writer_acq_1_no_pred}|
    /* Change the counters to WRITE as we have the lock now.*/
    set_counters_to_WRITE();|\label{line:acquire_cacc2_1}|
    Put(ACQUIRE_START, $p$, STATUS); Flush($p$); } }
\end{lstlisting}

\subsubsection{Writer Lock Release: Level 1 (Listing~\ref{lst:writer_release_1})}
\label{sec:writer_release_1}

% Listing~\ref{lst:writer_release_1} illustrates the protocol. 

\textbf{\textsf{Intuition:}}
$p$ first checks if it has reached $T_{W}$ and if there is a
successor waiting at level~1. If any case is true, it passes the lock
to the readers and notifies any successor that it must acquire the lock from
them. Otherwise, the lock is handed over to the successor.

\noindent
\textbf{\textsf{Details:}}
First, if $T_{W}$ is reached, $p$ passes the lock to the readers by resetting
the counters (Line~\ref{line:release_reset_1}).  Then, if it has no successor,
it similarly enables the readers to enter the CS
(Line~\ref{line:release_reset2_1}).  Later, $p$ appropriately modifies the tail
of the DQ and verifies if there is a new successor
(Line~\ref{line:release_cas_1}).  If necessary, it passes the lock to the
successor with a \texttt{Put} (line \ref{line:release_put_1}) and
simultaneously (using \texttt{next\_stat}) notifies it about a possible lock
mode change. 

% Then, it sets successor's status is set to \emph{MODE\_CHANGE} as a
% notification that the lock mode has changed.  Finally, a local flag is set to
% indicate that the counters have been reset. 

% the lock cannot be simply released. The process has to check the tail of the
% queue for new arrivals with \texttt{CAS} (line \ref{line:release_cas_1}) like
% in the MCS lock release. Being sure whether it has a successor or not, it
% will either pass the lock or successfully return. In any case, the process
% releases the lock to the readers (line \ref{line:release_reset2_1}) if that
% has not already happened.

% \begin{lstlisting}[float=h,caption=Function that resets one counter., label=lst:reset_counter]
% void reset_counter(int cnt_rank) {
%   int64_t arr_cnt, dep_cnt;
%   int64_t sub_arr_cnt, sub_dep_cnt;
% 
%   /* We first get the current values of the counters */
%   arr_cnt = Get(cnt_rank, ARRIVE);
%   dep_cnt = Get(cnt_rank, DEPART);
%   Flush(cnt_rank);
% 
%   /* We prepare the values that will be subtracted from the counters */
%   sub_arr_cnt = -dep_cnt;
%   sub_dep_cnt = -dep_cnt;
% 
%   /* We need to make sure that we reset WRITE if it was set */
%   if(arr_cnt >= WRITE) 
%   {
%     sub_arr_cnt -= WRITE;
%   }
%   /* Subtract the values from the current counters */
%   Accumulate(sub_arr_cnt, cnt_rank, ARRIVE, SUM);
%   Accumulate(sub_dep_cnt, cnt_rank, DEPART, SUM);
%   Flush(cnt_rank);
% }
% \end{lstlisting}

\begin{lstlisting}[float=h,caption=Releasing an RMA-RW lock by a writer; level
1., label=lst:writer_release_1]
void writer-release<1>(){
  bool counters_reset = false;
  /* Get the count of consecutive lock acquires (level 1).*/
  int next_stat = Get($p$, STATUS); Flush($p$);
  if(++next_stat == $T_{W}$) { /* Pass the lock to the readers.*/
    reset_counters();|\label{line:release_reset_1}|/* See Listing |\ref{lst:manipulate_counters}|.*/
    next_stat = MODE_CHANGE; counters_reset = true;
  }
  int succ = Get($p$, NEXT); Flush($p$);
  if(succ == $\emptyset$) { /* No known successor.*/
    if(!counters_reset) { /* Pass the lock to the readers.*/ 
      reset_counters(); next_stat = MODE_CHANGE;|\label{line:release_reset2_1}|/* Listing |\ref{lst:manipulate_counters}|.*/
    }
    /* Check if some process has already entered the DQ.*/
    int curr_rank = CAS($\emptyset$, $p$, tail_rank[1,$e(p,1)$], TAIL);
    Flush(tail_rank[1,$e(p,1)$]);
    if($p$ == curr_rank) { return; } /* No successor...*/ |\label{line:release_cas_1}|
    do { /* Wait until the successor makes itself visible.*/
      succ = Get($p$, NEXT); Flush($p$);
    } while (succ == $\emptyset$); 
  }
  /* Pass the lock to the successor.*/
  Put(next_stat, succ, STATUS); Flush(succ);|\label{line:release_put_1}| }
\end{lstlisting}

\subsubsection{Reader Lock Acquire (Listing~\ref{lst:reader_acquire})}
\label{sec:reader_acquire}

\textbf{\textsf{Intuition:}}
Here, $p$ first spin waits if there is an active writer or if
$p$'s arrival made its associated counter $c(p)$ exceed $T_{R}$. Then, it can
enter the CS. If $c(p) = T_{R}$, then $p$ resets DC.

\noindent
\textbf{\textsf{Details:}}
In the first part, $p$ may spin wait on a boolean \texttt{barrier}
variable (Line~\ref{line:reader_if}), waiting to get the lock from a writer.
Then, $p$ atomically increments its associated counter and checks whether the
count is below $T_{R}$. If yes, the lock mode is \texttt{READ} and $p$ enters
the CS. Otherwise,  either the lock mode is \texttt{WRITE} or $T_{R}$ is
reached. In case of the latter, $p$ checks if there are any waiting writers
(Line~\ref{line:reader_acq_check_writers}).  If there are none, $p$ resets the
DC (Line~\ref{line:reader_reset}) and re-attempts to acquire the lock. If there is
a writer, $p$ sets the local barrier and waits for DC to be reset by the
writer.

\begin{lstlisting}[float=h,caption=Acquiring an RMA-RW lock by a reader.,
label=lst:reader_acquire]
void reader-acquire() {
  bool done = false; bool barrier = false;
  while(!done) {
    int curr_stat = 0;
    if(barrier) { |\label{line:reader_if}|
      do {
        curr_stat = Get($c(p)$, ARRIVE); Flush($c(p)$);
      } while(curr_stat >= $T_{R}$);
    }

    /* Increment the arrival counter.*/
    curr_stat = FAO(1, $c(p)$, ARRIVE, SUM); Flush($c(p)$);
    if(curr_stat >= $T_{R}$) { /* $T_{R}$ has been reached...*/
      barrier = true; 
      if(curr_stat == $T_{R}$) {/* We are the first to reach $T_{R}$.*/ 
        /* Pass the lock to the writers if there are any.*/
        int curr_tail = Get(tail_rank[1,$e(p,1)$], TAIL);|\label{line:reader_acq_check_writers}|
        Flush(tail_rank[1,$e(p,1)$]); 
        if(curr_tail == $\emptyset$) { /* There are no waiting writers.*/
          reset_counter($c(p)$); barrier = false;|\label{line:reader_reset}|/* Listing |\ref{lst:manipulate_counters}|.*/
        } 
      }
      /* Back off and try again.*/
      Accumulate(-1, $c(p)$, ARRIVE, SUM); Flush($c(p)$); 
    } } }
\end{lstlisting}

\subsubsection{Reader Lock Release (Listing~\ref{lst:reader_release})}
\label{sec:reader_release}

Releasing a reader lock only involves incrementing the departing
reader counter. 

\begin{lstlisting}[caption=Releasing an RMA-RW reader lock., label=lst:reader_release]
void reader-release() {
  Accumulate(1, $c(p)$, DEPART, SUM); Flush($c(p)$); }
\end{lstlisting}

\subsection{Example}
\label{sec:example}

Consider the scenario from Figure~\ref{fig:structures}. Here, there are three
machine levels, 12 readers, and 12 writers ($F_W = 0.5$). 

\macb{Writer Acquire }
Assume a new writer $W_x$ running on a node related to DQ$_{3.1}$ attempts to
acquire RMA-RW (Figure~\ref{fig:structures}, Part~5). First, it enters DQ$_{3.1}$
(Listing~\ref{lst:writer_acquire_n}). 
%
% For this, it uses the \texttt{tail\_rank} and $e(\cdot,\cdot)$ mappings.
%
As this queue is not empty, $W_x$ spins locally (Lines~10-12)
until its predecessor $W_9$ modifies $W_x$'s \texttt{STATUS}. 
Now, if $W_9$ has not yet reached $T_{L,3}$, $W_x$ gets
the lock and immediately proceeds to the CS (Lines~15-19).  Otherwise, it
attempts to move to level~2 by updating its \texttt{STATUS} (Line~22) and
calling \texttt{writer-acquire<$i-1$>()}.
Thus, it enters DQ$_{2.1}$ and takes the same steps as in DQ$_{3.1}$: it spins
locally until $W_4$ changes its \texttt{STATUS} and it either directly
enters the CS or it proceeds up to level~1.  Assuming the latter, $W_x$ enters
DQ$_{1.1}$ and waits for $W_1$ to change its \texttt{STATUS}
(Listing~\ref{lst:writer_acquire_1}, Lines~10-12).
If \texttt{STATUS} is different from \texttt{MODE\_CHANGE} (Line~17),
$W_x$ can enter the CS. Otherwise, the lock was handed over to the readers and
$W_x$ calls \texttt{set\_counters\_to\_WRITE()} to change all physical counters
to the \texttt{WRITE} mode (Line~15), which blocks new incoming readers.
At some point, the readers reach the $T_{R}$ threshold and hand the lock over
to $W_x$. 

\macb{Writer Release }
Assume writer $W_x$ occupies the CS and starts to release RMA-RW
(Figure~\ref{fig:structures}, Part~6). It begins with level~3
(Listing~\ref{lst:writer_release_n}). Here, it first checks if it has a
successor in DQ$_{3.1}$ and if $T_{L,3}$ is not yet reached (Line~5).  Its
successor is $W_{10}$ and assume that the latter condition is true. Then, $W_x$
passes the lock to $W_{10}$ by updating its \texttt{STATUS} so that it contains
the number of lock acquires within the given element. If $T_{L,3}$ is reached,
$W_x$ releases the lock at level~2 (Line~12). Here, it repeats all the above
steps (its successor is $W_6$) and then starts to release the lock at level~1
(Listing~\ref{lst:writer_release_1}). Here it hands the lock over to the
readers if $T_W$ is reached (Lines~5-8). Finally, it notifies its successors at
each level ($N$ to~2) to acquire the lock at the parent level
(Listing~\ref{lst:writer_release_n}, Line~23).

\macb{Reader Acquire }
A reader $R_x$ that attempts to acquire RMA-RW first increments $c(R_x)$
(Listing~\ref{lst:reader_acquire}, Line~12) and checks if $T_R$ is reached (in
the first attempt Lines~6-8 are skipped). If yes, it sets \texttt{barrier}
(Line~14), backs off (Line~24), and reattempts to acquire the lock. In
addition, if $R_x$ is the first process to reach $T_R$, it also checks if there
are any waiting writers (Lines~15-21). If not, it resets $c(R_x)$ and sets
\texttt{barrier} to \texttt{false} so that it can enter the CS even if $T_R$
was reached. Then, it reexecutes the main loop (Line~3); this time it may enter
the loop in Lines~6-8 as the lock was handed over to a writer (if $T_R$ was
reached). In that case, $R_x$ waits until its $c(R_x)$ is reset
(Listing~\ref{lst:reader_acquire}, Lines~6-8). 

\macb{Reader Release }
This is a straightforward scenario in which $R_x$ only increments \texttt{DEPART}
at $c(R_x)$.

% \subsection{Implementation Details}
% 
% In Listing \ref{lst:constants} we present all the constants that we use and in
% Listing \ref{lst:queue_node} we provide the declarations of the three utilized
% data structures.
% 
% \begin{lstlisting}[caption=Declaration of the used constants., label=lst:constants]
% /* Constants used for passing information between the writers */
% enum {COHORT_START=0, CHANGE_MODE = INT64_MAX -1, ACQUIRE_PARENT = INT64_MAX - 2, WAIT = INT64_MAX};
% /* Constants used for solving the conflict between readers and writers */
% enum {WRITE = INT64_MAX / 2, START_READ = 0};
% /* Offsets for the memory window */
% enum {NEXT = 0,  STATUS = 1, LOCK_NODE = 2, TAIL = 3, THRESHOLD = 4};
% /* Offsets for the counter array */
% enum {ARRIVE = 0, DEPART = 1};
% \end{lstlisting}
% %
% %\begin{lstlisting}[caption=Declaration of the three utilized data structures, label=lst:queue_node]
% %int64_t queueNode[2];
% %int64_t hierarchyNode[4];
% %int64_t counter[2];
% %\end{lstlisting}

\subsection{RMA-RW vs. RMA-MCS}
\label{sec:topo_mcs}

We also outline the design of RMA-MCS.
RMA-MCS consists of DQs and DT but not DC.  $T_R$ and $T_W$ are excluded as the
are no readers.  Similarly, $T_{L,1}$ is not applicable because there is no
need to hand the lock to readers.
The acquire/release protocols are similar to the ones in
Listings~\ref{lst:writer_acquire_n} and~\ref{lst:writer_release_n} for any $i
\in \{1, ..., N\}$.

\begin{figure*}
\centering
%\vspace{-1.3em}
 \subfloat[Latency (LB).]{
  \includegraphics[width=0.185\textwidth]{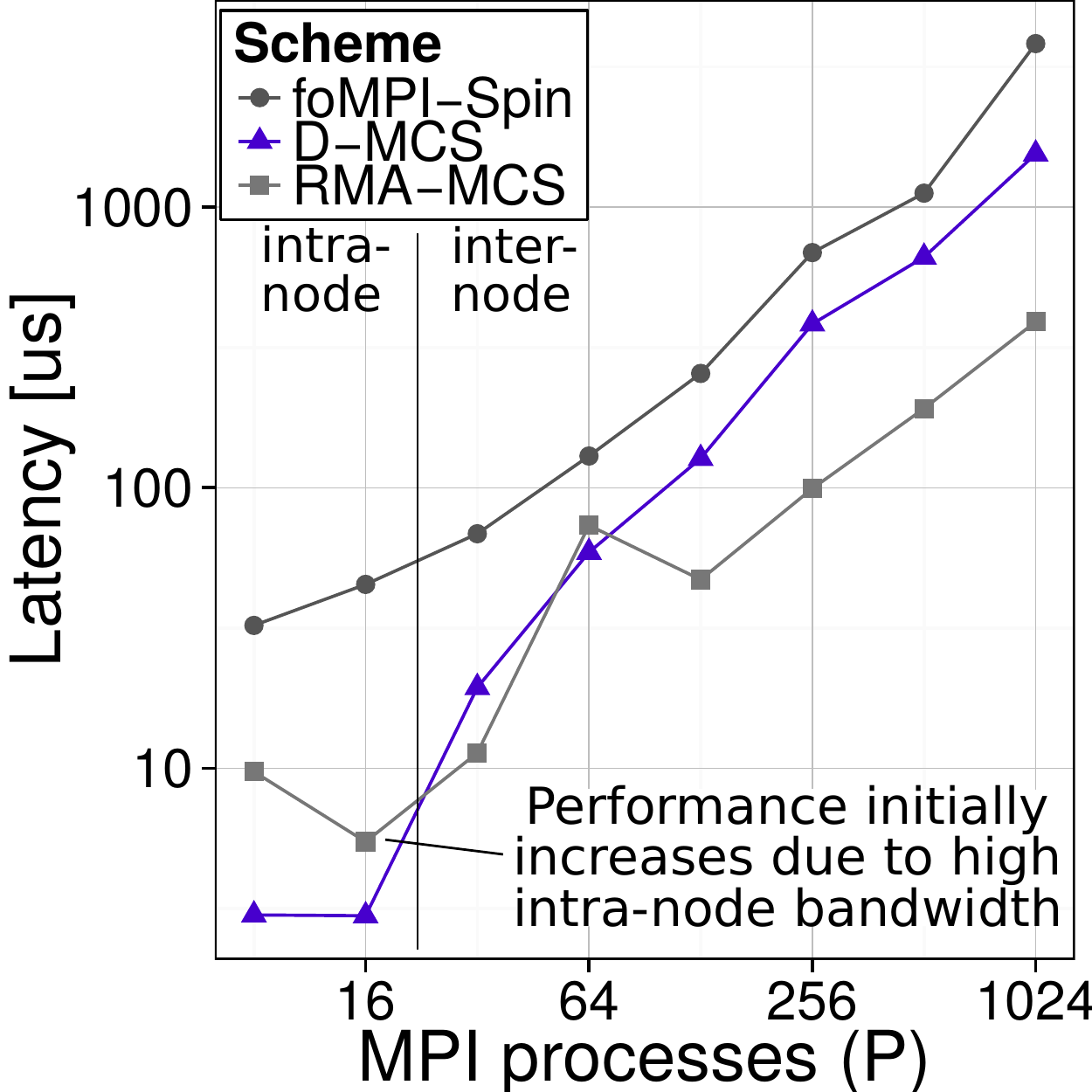}
  \label{fig:mcs_fompi_latency_comp}
 }\hfill
 \subfloat[Throughput (ECSB).]{
  \includegraphics[width=0.185\textwidth]{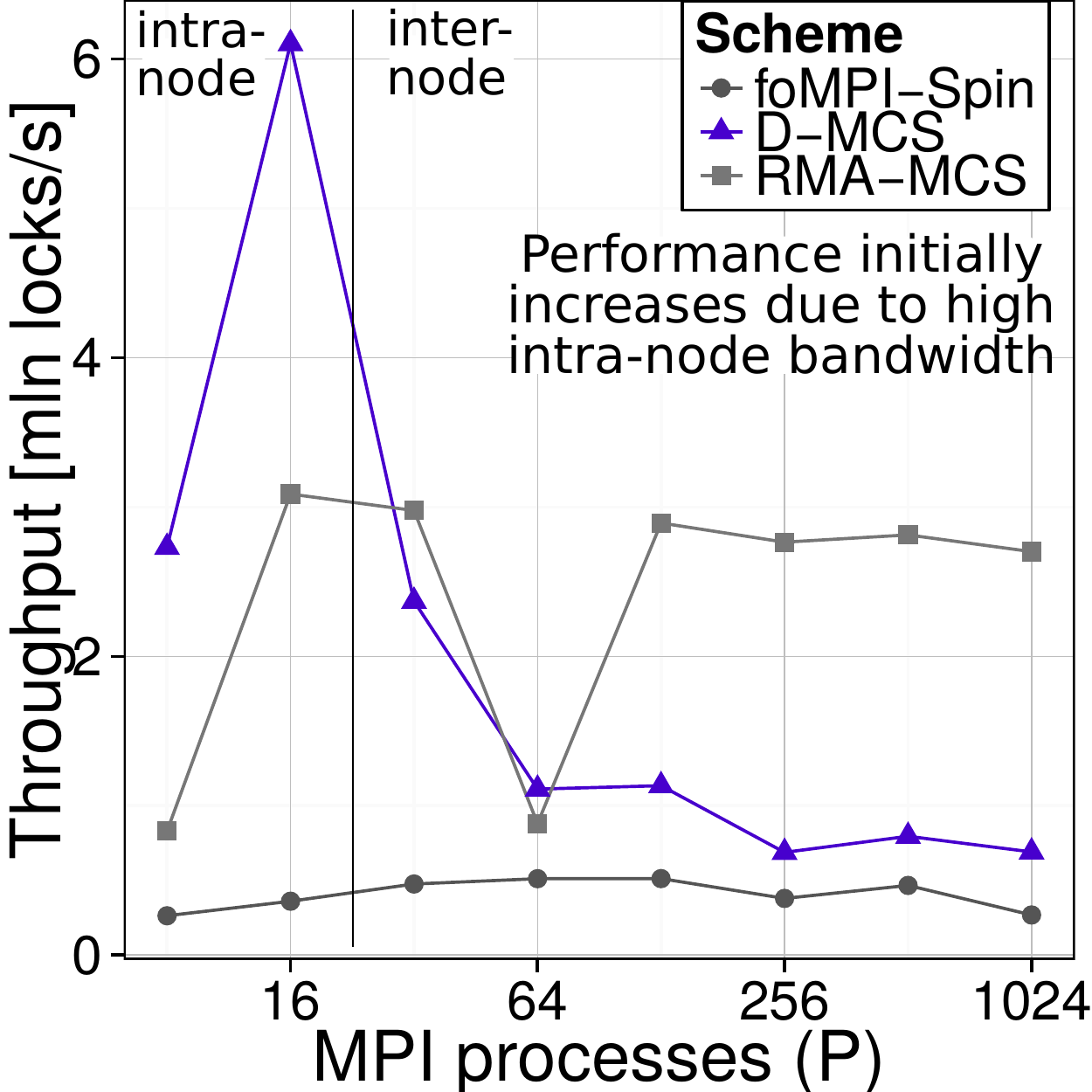}
  \label{fig:queue_ecsb_perf}
 }\hfill
 \subfloat[Throughput (SOB).]{
  \includegraphics[width=0.185\textwidth]{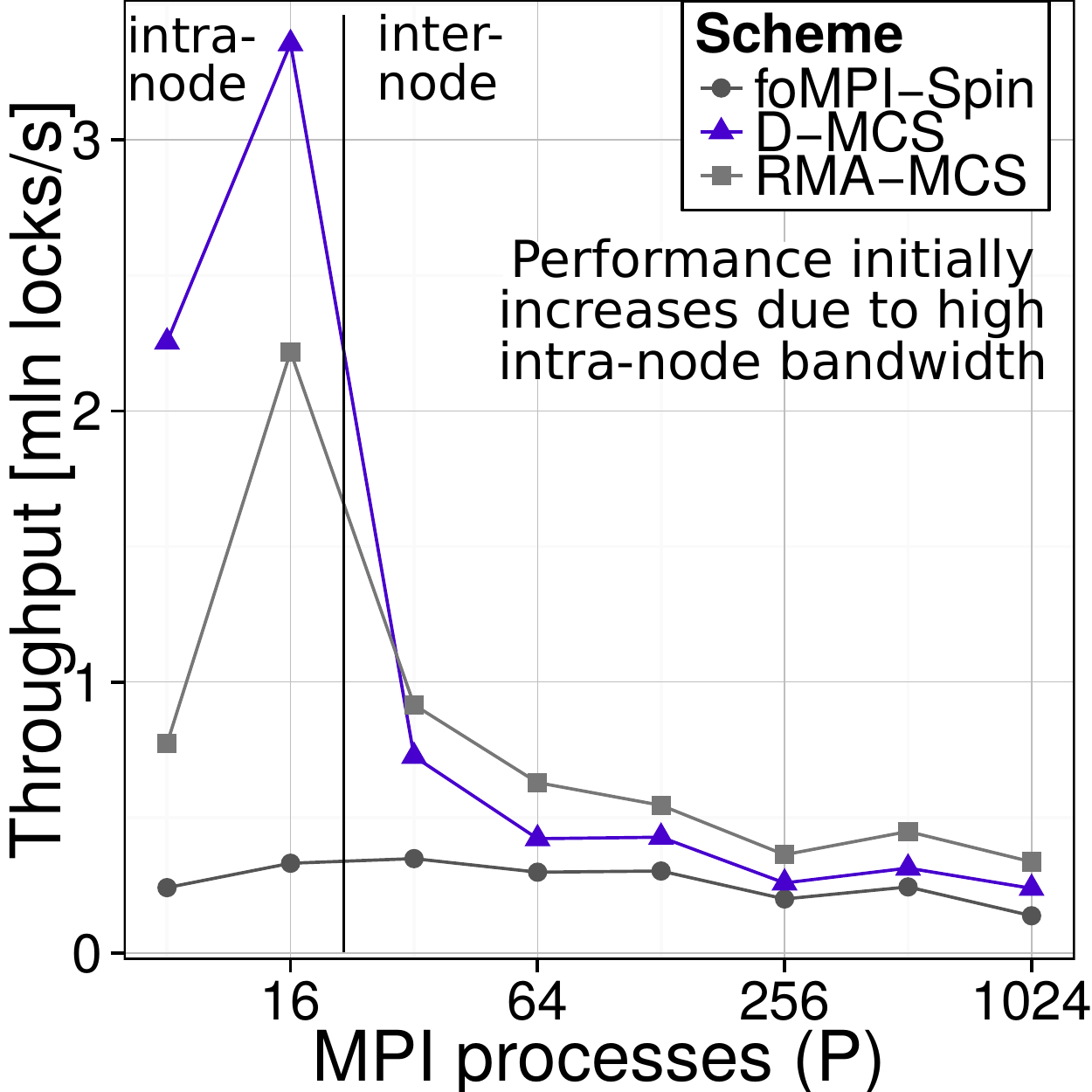}
  \label{fig:queue_sob_perf}
 }\hfill
 \subfloat[Throughput (WCSB).]{
  \includegraphics[width=0.185\textwidth]{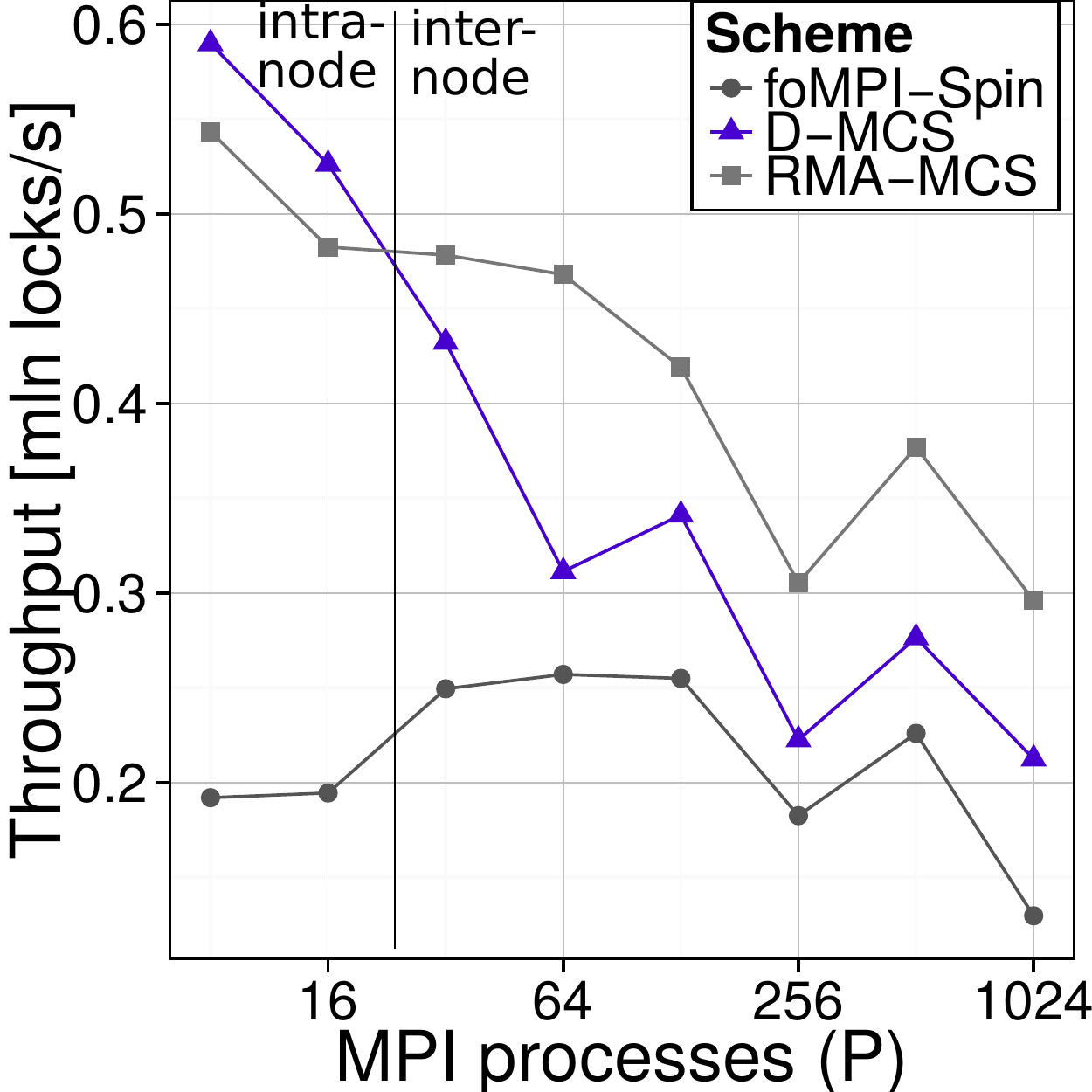}
  \label{fig:queue_wcsb_perf}
 }\hfill
 \subfloat[Throughput (WARB).]{
  \includegraphics[width=0.185\textwidth]{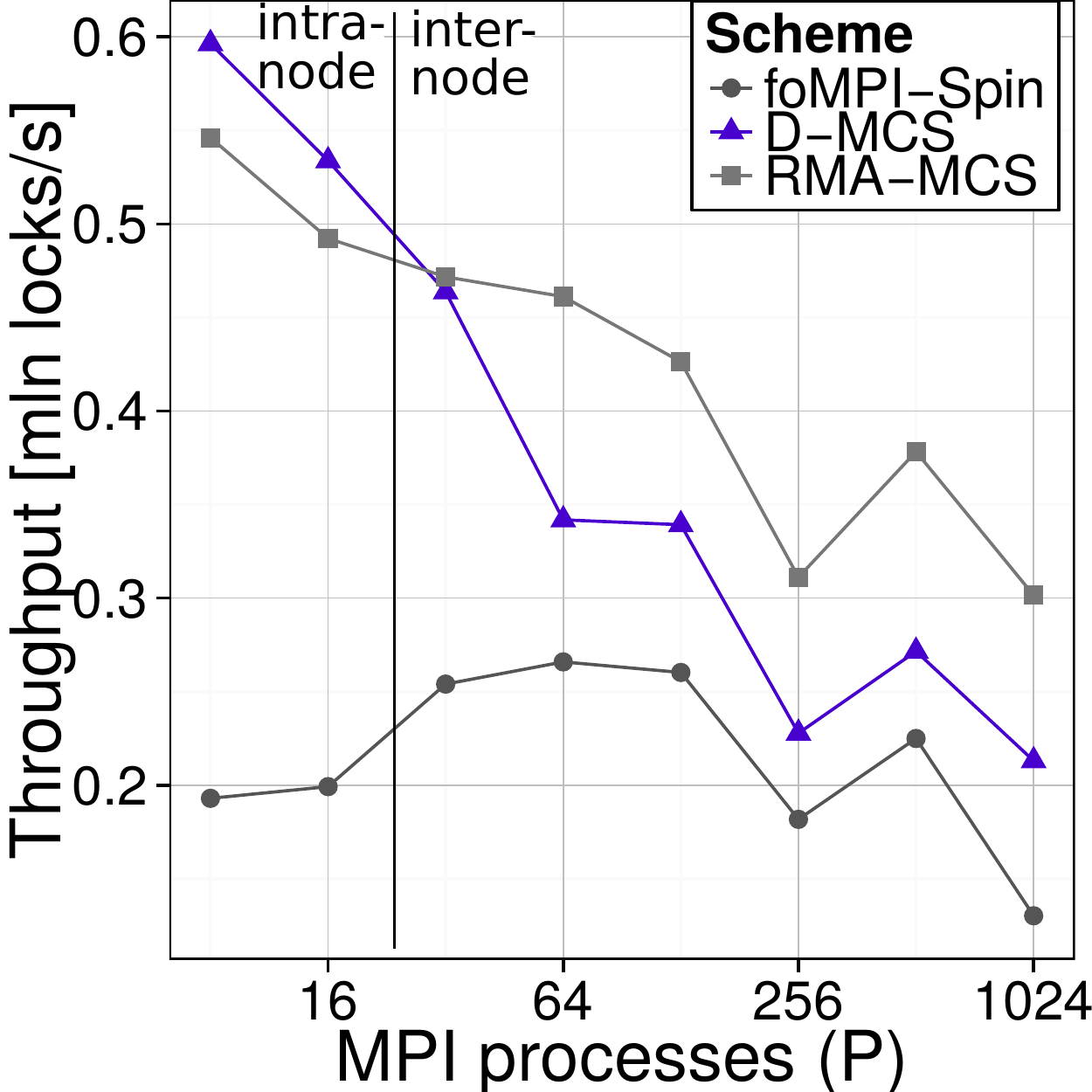}
  \label{fig:queue_warb_perf}
 }
%\vspace{-0.5em}
\caption{(\cref{sec:eval_mcs_lock}) Performance analysis of RMA-MCS and comparison to the state-of-the-art.}
%\vspace{-1.5em}
\label{fig:rma_mcs_perf}
\end{figure*}

% \section{Correctness Analysis}
\section{CORRECTNESS ANALYSIS}

We now discuss how RMA-RW ensures three fundamental correctness
properties: mutual exclusion (ME), deadlock freedom (DF), 
and starvation freedom (SF)~\cite{Herlihy:2008:AMP:1734069}. 
At the end of this section, we show how we use model checking to 
verify the design.

% In the first step, we show that it guarantees mutual exclusion. In the
% second, we illustrate that the protocols are deadlock-free. 

\subsection{Mutual Exclusion}

% \begin{thm}
% RMA-RW provides mutual exclusion.
% \end{thm}

% \begin{proof}
% %
% There are two possible ways how mutual exclusion is violated: two writers
% access a critical section concurrently or a reader and a writer. 
% 
% \textbf{\textsf{Writer vs. Writer }}
% %
% We have to distinguish between writers that work on the same NUMA domain or
% on different ones. Let us first consider the case where they sit on the same
% which means that they already operate on the same \texttt{TAIL}. Thus, they
% can only acquire the lock at the same time if both writers do not see any
% predecessor. But this is not possible because we use the atomic function
% \texttt{FAO} for modifying the \texttt{TAIL} field that guarantees that one
% fetches the value before the other does. On the other hand, two writers on
% different domains have a common parent node where they have to fight for the
% lock. At this level, we can apply the same argument as before since the MCS
% lock is acquired the same on each level of the tree.
% 
% \textbf{\textsf{Reader vs. Writer}}
% %
% A reader and a writer can be active at the same time if the mode of the
% system is on \texttt{READ} and about to change to \texttt{WRITE}. This is
% because the reader cannot change the mode themselves and as a consequence
% cannot acquire a lock while a writer is active. However, a writer can alter
% the mode to \texttt{WRITE} during a reader is active. But this is no problem
% since a writer checks each counter again for active readers after changing
% all of them.
% %
% \end{proof}

% \begin{proof}
%
ME is violated if two writers or a reader and a writer enter
the CS concurrently. We now discuss both cases.

\textbf{\textsf{Writer \& Writer: }}
We distinguish between writers that are in the same DQ (case~A) or in different
ones (case~B). In case~A, they operate on the same \texttt{TAIL}. Thus, they
could only violate ME if both writers do not see any predecessor. This is
prevented by using \texttt{FAO} for atomically modifying \texttt{TAIL}.
In case~B, two writers competing in different DQs have a common DQ in DT where
they or their predecessor compete for the lock. Similarly as above, the MCS
lock must be acquired at each DT level. If a predecessor has to
compete for the lock, a writer waits until he gets notified by its predecessor
and thus does not interfere in the lock acquiring process.

\textbf{\textsf{Reader \& Writer: }}
A reader and a writer can be active at the same time if the lock mode 
is \texttt{READ} and about to change to \texttt{WRITE}. This is because the
reader on its own cannot change the mode and as a consequence cannot acquire a
lock while a writer is active. However, a writer can alter the mode to
\texttt{WRITE} while a reader is active. This is prevented by a writer
that checks each counter again for active readers after changing all of them.
%
% \end{proof}

\subsection{Deadlock Freedom}

% \begin{thm}
% RMA-RW provides deadlock freedom.
% \end{thm}

% \begin{proof}
%
Here, we also differentiate two base cases: two writers deadlock 
or a reader and a writer deadlock. 

% \maciej{Not sure... Maybe a better approach is to see where in the code
% a process can get stuck, and show this cannot happen}.

\textbf{\textsf{Writer \& Writer}}
%
%The only way that two writers deadlock is if each one becomes the predecessor
%of the other. 
The only way how writers deadlock is if there is a cycle in a queue. For two writers 
it means that one becomes the predecessor of the other. Therefore, both wait on 
the other to get notified. This cannot happen as the processes use an atomic 
\texttt{FAO} to obtain their predecessor. As explained, this function is atomic and 
thus we can order the uses of \texttt{FAO} in a timeline. This contradicts that 
the writers have a cycle in their waiting queue. 
%
%\maciej{I don't think it really exhausts all the cases?  What if we have 3
%writers? or N writers?}

\textbf{\textsf{Reader \& Writer}}
A reader may deadlock after $T_{R}$ is reached (case~A) or the mode goes into
\texttt{WRITE} (case~B). In case~A, either there is no writer active and the
reader resets the DC or a writer is waiting and a reader backs off. Thus, the
writer changes the mode to \texttt{WRITE} after all readers back off which is
done in a finite time.  As writers do not deadlock and the last writer changes
the mode back to \texttt{READ}, no reader will deadlock in case~B either.
%
% \end{proof}

\subsection{Starvation Freedom}

% \begin{thm}
% RMA-RW provides starvation freedom.
% \end{thm}

% \begin{proof}
%
Finally, we show that no writer or reader can starve. 

\textbf{\textsf{Writers}}
A writer may starve while other writers or readers are active. 
We prevent it with different thresholds. First, there is $T_{L,i}$
at each DT level~$i$. After reaching $T_{L,i}$, writers in
one of the associated DQs at $i$ release the lock to the next DQ at the same
level. Thus, we only need to show that one DQ is starvation-free which is
already provided by the underlying MCS queue lock design. Yet,
there is the $T_{R}$ threshold that regulates the number of lock acquires by readers for
one counter before the readers associated to the counter back off. We already
showed that the readers make progress. Thus, at some point, all counters have
reached $T_R$ and a writer changes the mode to \texttt{WRITE}.

\textbf{\textsf{Readers}}
There are two ways how readers could starve. First, other readers are active
while processes associated with a certain counter back off to let writers
  acquire the lock. However, there is the $T_R$ threshold for each counter
  after which the readers associated with this counter back off. Thus,
  eventually, all readers wait on the writers to take over.
  %
  %~\maciej{This doesn't show that readers won't starve, no? They are still stuck, no?}. 
  This leads us to the second case where the writers have the lock and do not pass it 
  to the waiting readers. This is not possible since there is the $T_{L,i}$ threshold at 
  each level of the writer hierarchy and at most after $T_{W} = \prod_{i=1}^{N} T_{L,i}$ lock passings
  between writers the lock goes to readers; we have also already illustrated
  that the writers will make progress until this threshold is reached.
  
  \subsection{Model Checking}
 
To confirm that RMA-RW provides the desired correctness properties, we also
conduct model checking with SPIN~\cite{Holzmann:1997:MCS} (v6.4.5), a software
tool for the formal verification of multi-threaded codes.  The input to SPIN is
constructed in PROMELA, a verification modeling language that allows for the
dynamic creation of concurrent processes to model, for example, distributed
systems.
We evaluate RMA-RW for up to $N \in \{1, ..., 4\}$ and a maximum of 256
processes. The machine elements on each level of the simulated system have the
same number of children. Thus, for $N = 3$ and four subelements per machine
element, the system would consist of $4^{3}$ processes.  Each process is
defined randomly either as a reader or a writer at the beginning and after that, it
tries to acquire the lock 20 times. We choose this value as it generates a
feasible number of cases that SPIN has to check even for a high count of
processes. During the execution of a test, we use a designated process that
verifies that either only one writer or multiple readers hold a lock.  All the
tests confirm mutual exclusion and deadlock freedom.

\section{EVALUATION}

% We now illustrate performance advantages of RMA-MCS
% (\cref{sec:eval_mcs_lock}) and RMA-RW (\cref{sec:eval_rw_lock}) over
% state-of-the-art distributed locks from the foMPI implementation of MPI-3
% RMA~\cite{fompi-paper}.

We now illustrate performance advantages of RMA-MCS and RMA-RW over
state-of-the-art distributed locks from the foMPI implementation of MPI-3
RMA~\cite{fompi-paper}.

% We begin this chapter by explaining the setup that we use for our experiments
% (\cref{sec:settings}).  Afterwards, we show the results of the evaluation of
% the RMA-MCS (\cref{sec:eval_mcs_lock}) and then of the RMA-RW lock
% (\cref{sec:eval_rw_lock}).

\textbf{\textsf{Comparison Targets }}
We consider D-MCS and both foMPI locking schemes: a simple spin-lock
(\texttt{foMPI-Spin}) that enables mutual exclusion, and an RW lock
(\texttt{foMPI-RW}) that provides both shared and exclusive accesses to the CS.

% We use an implementation of a state-of-the-art RW lock of a MPI
% implementation (foMPI) \cite{fompi-paper} as a comparison. foMPI has two
% locking schemes that we utilize: 
% 
% \begin{itemize} % \item A simple spin-lock (\emph{foMPI-Spin}) that allows
% mutual exclusion % \item A RW lock algorithm (\emph{foMPI-RW}) that provides
% shared and exclusive access \end{itemize}
% 
% We did not compare to other state-of-the-art NUMA-aware locks because they do
% not approach distributed locks and thus are orthogonal to our scheme. Their
% solutions however could be incorporated into our approach to make it even
% more scalable.
% 
% We first present the results of the MCS lock compared to the foMPI-Spin. We
% thereafter show how we set the parameters for an optimized version of the
% RMA-RW lock. At the end, we demonstrate how well the RMA-RW lock performs
% compared to the RMA-MCS and foMPI-RW lock. 

\textbf{\textsf{Selection of Benchmarks }}
We conduct six series of experiments. The latency benchmark (LB) measures the
latency of both acquiring and releasing a lock; an important performance
metric in workloads such as real-time queries. Four other analyses obtain
throughput under varying conditions and parameters. The empty-critical-section
benchmark (ECSB) derives the throughput of acquiring an empty lock with no
workload in the CS. The single-operation benchmark (SOB) measures the
throughput of acquiring a lock with only one single operation (one memory
access) in the CS; it represents irregular parallel workloads such as graph
processing with vertices protected by fine locks. Next, the
workload-critical-section benchmark (WCSB) covers variable workloads in the CS:
each process increments a shared counter and then spins for a random time
(1-4$\mu$s) to simulate local computation. The wait-after-release benchmark
(WARB) varies lock contention: after release, processes wait for a random time
(1-4$\mu$s) before the next acquire. The throughput experiments represent data-
and communication-intensive workloads. 
Finally, we integrate and evaluate the proposed locks with a distributed
hashtable (DHT) to cover real codes such as key-value stores.

\begin{figure*}
\centering
%\vspace{-1.3em}
 \subfloat[(\cref{sec:eval_t_dc}) $T_{DC}$ analysis, SOB, $F_W = 2\%$.]{
  \includegraphics[width=0.3\textwidth]{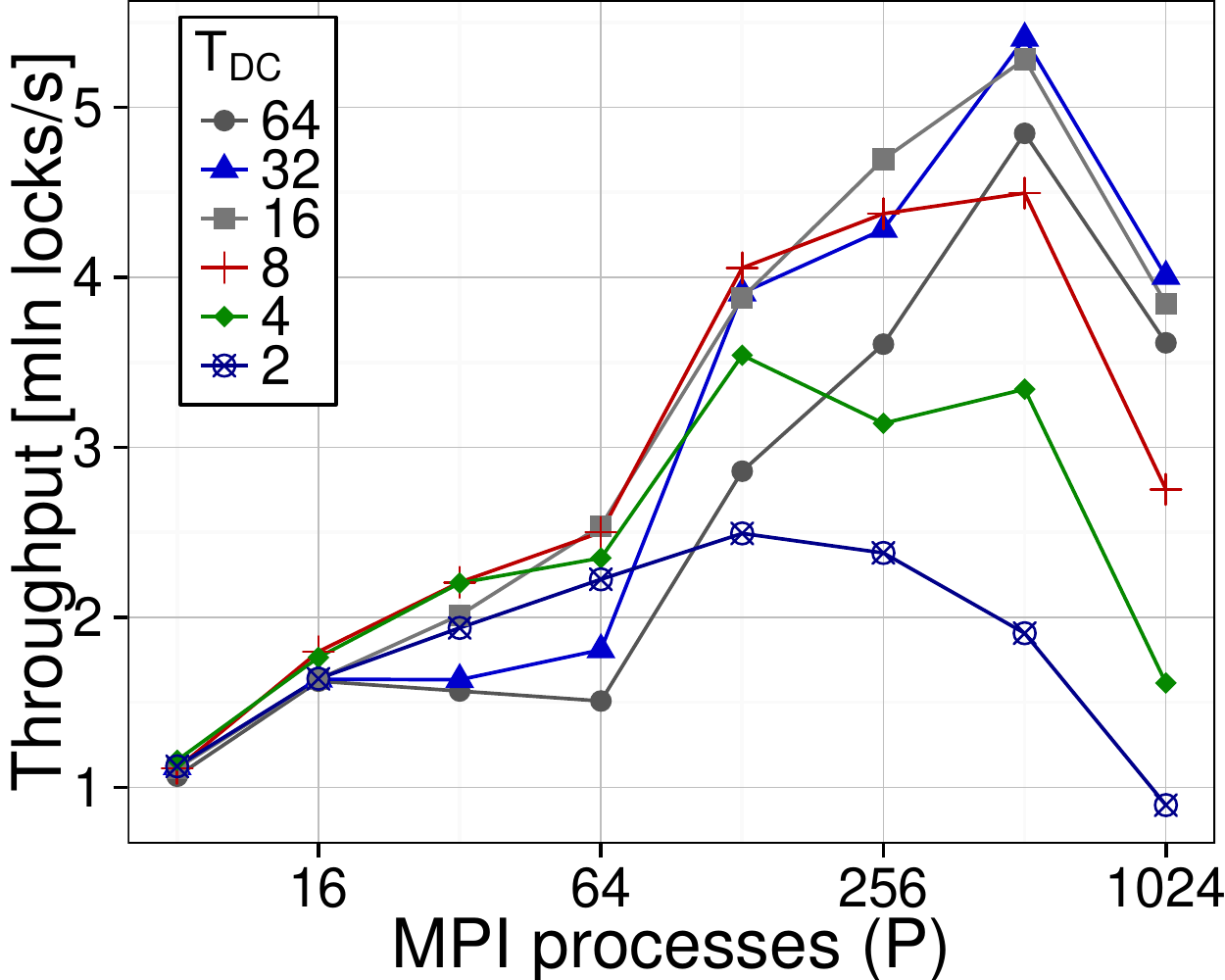}
  \label{fig:rma_rw_T_DC}
 }\hfill
 \subfloat[(\cref{sec:eval_t_li}) $\prod_{i=1}^{N} T_{L,i}$ analysis, SOB, $F_W = 25\%$.]{
  \includegraphics[width=0.3\textwidth]{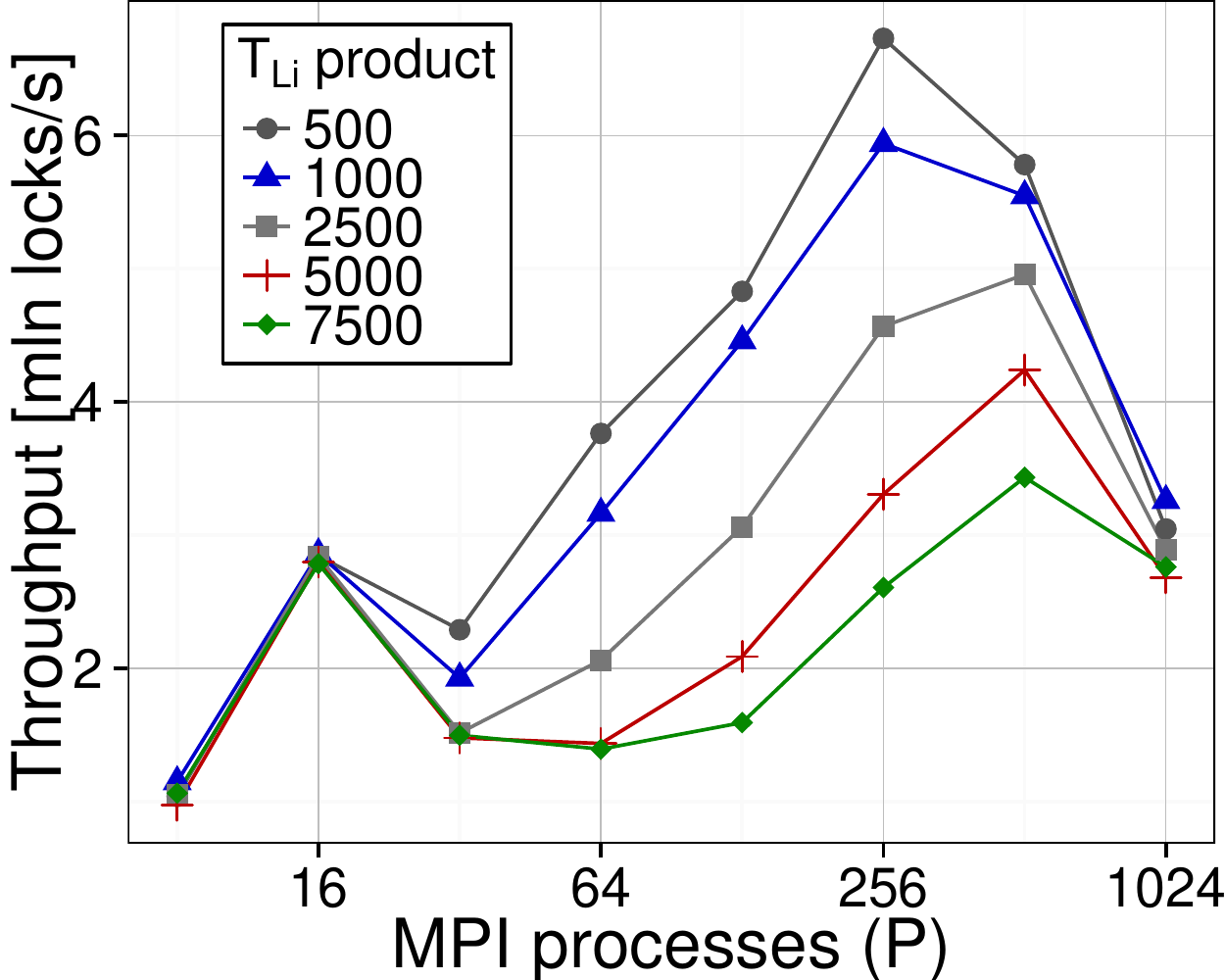}
  \label{fig:rma_rw_T_Li_prod}
 }\hfill
 \subfloat[(\cref{sec:eval_t_li}) $T_{L,i}$ analysis, SOB, $F_W = 25\%$.]{
  \includegraphics[width=0.3\textwidth]{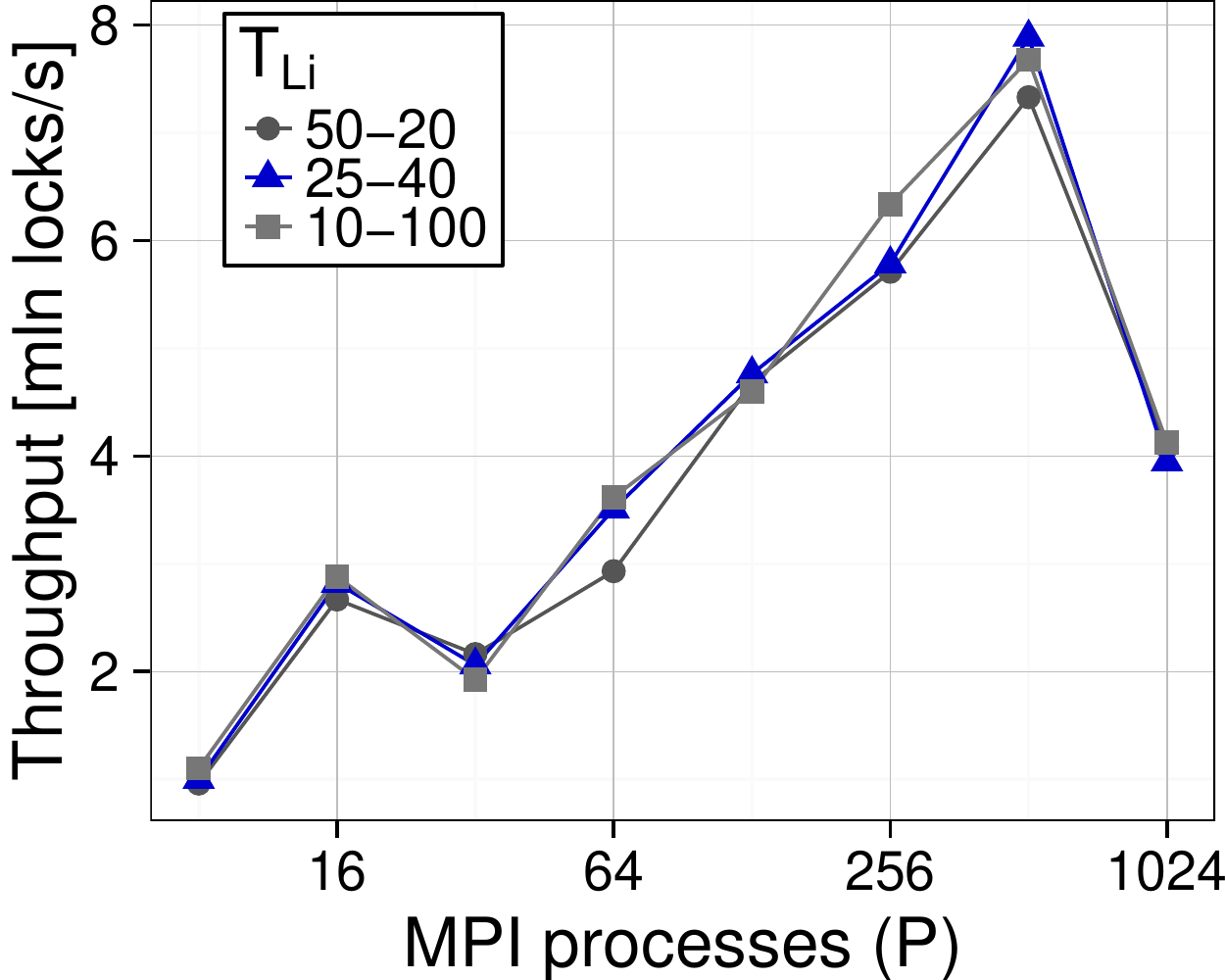}
  \label{fig:rma_rw_T_Li}
 }\\
 \subfloat[(\cref{sec:eval_t_li}) $T_{L,i}$ analysis, LB, $F_W = 25\%$.]{
  \includegraphics[width=0.3\textwidth]{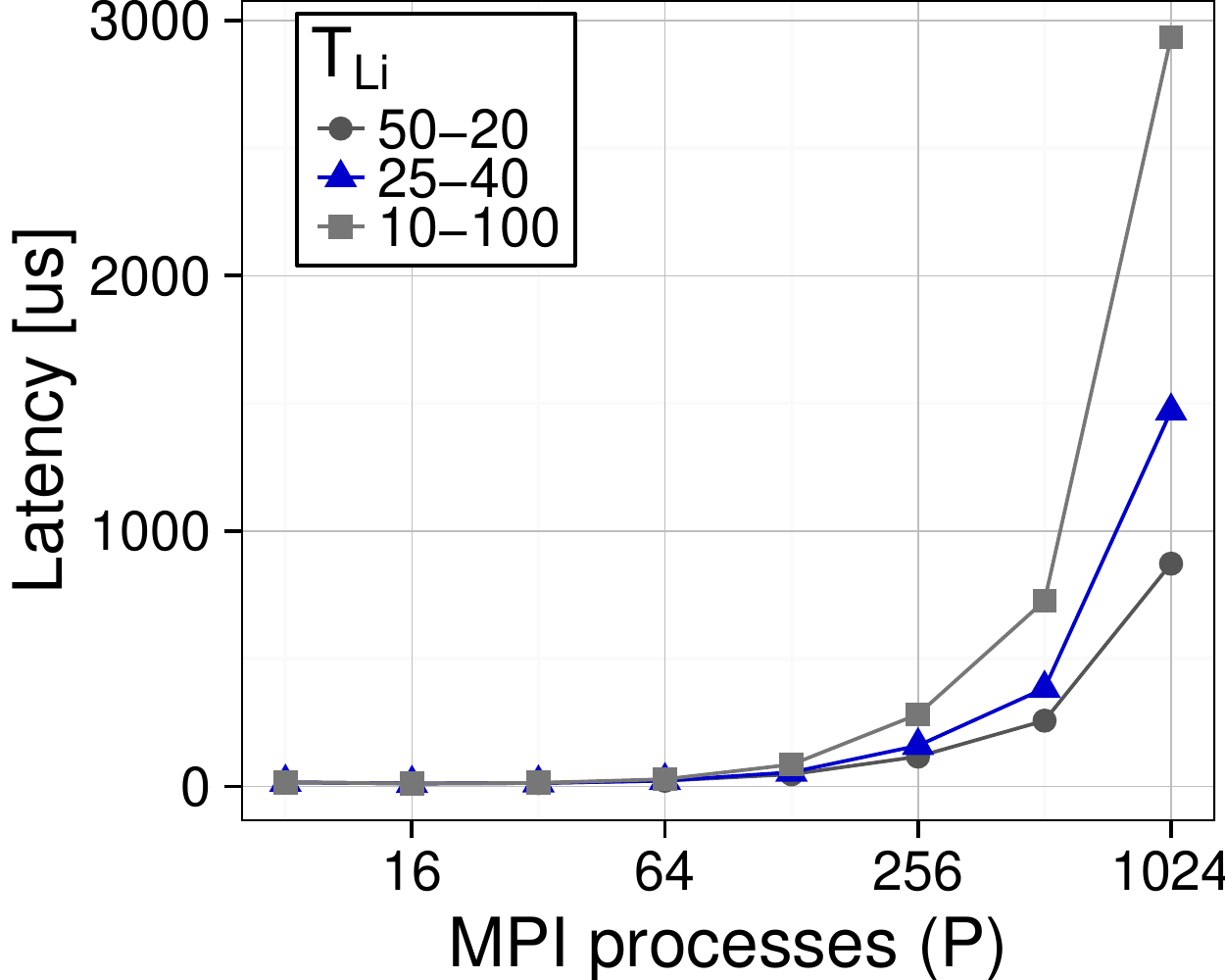}
  \label{fig:rma_rw_T_Li_LB}
 }\hfill
 \subfloat[(\cref{sec:eval_t_rw}) $T_{R}$ analysis, ECSB, $F_W=0.2\%$.]{
  \includegraphics[width=0.3\textwidth]{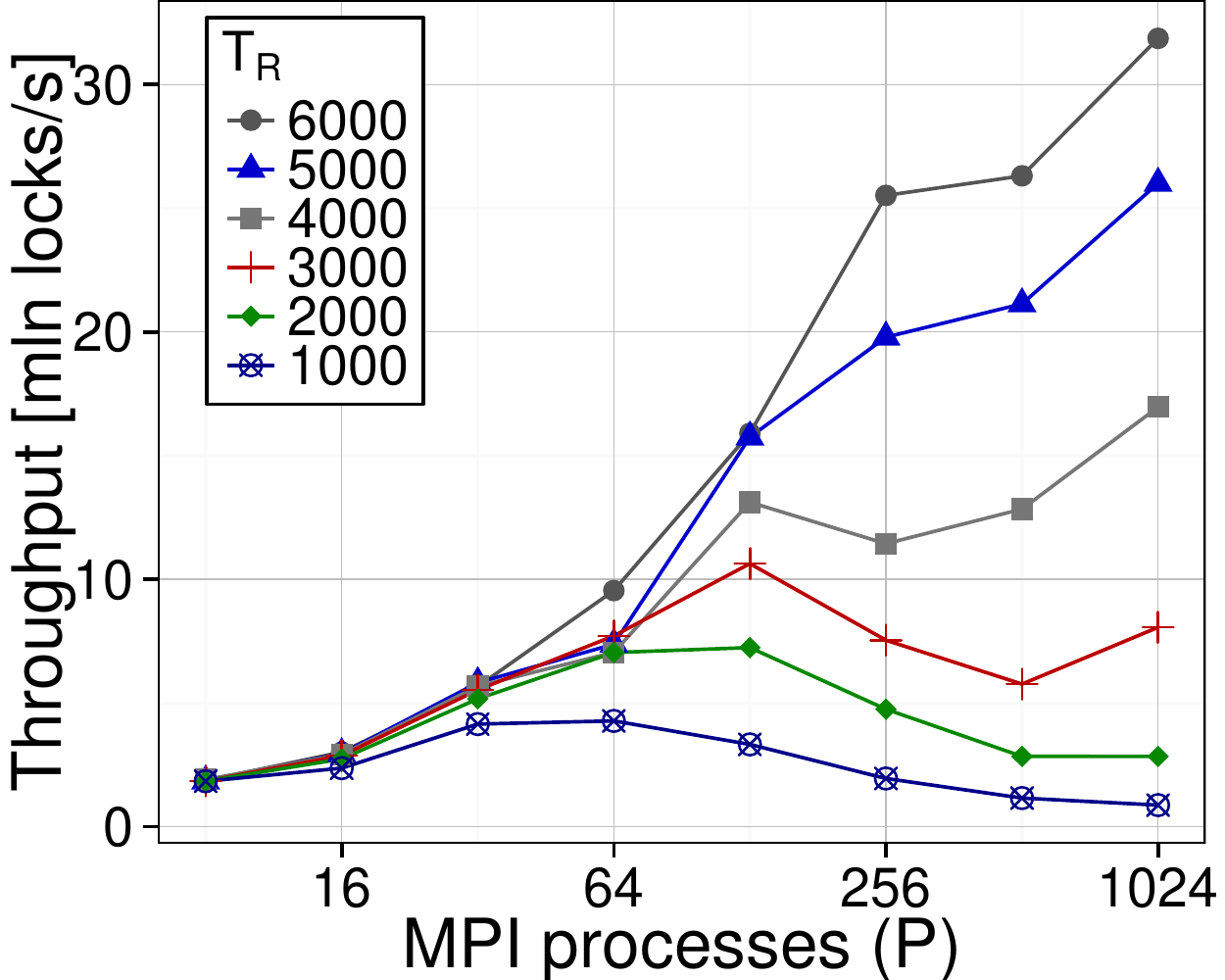}
  \label{fig:rma_rw_T_R_one_rate}
 }\hfill
 %
% \subfloat[(\cref{sec:eval_t_rw}) $T_{R}$ analysis, LB, $F_W \in \{2\%, 5\%\}$.]{
%  \includegraphics[width=0.31\textwidth]{rw_reader_t_latency_analysis_multiple_rates-eps-converted-to.pdf}
%  \label{fig:rma_rw_T_R_m_rates_LB}
% }\hfill
% %
 \subfloat[(\cref{sec:eval_t_rw}) $T_{R}$ analysis, ECSB, $F_W \in \{2\%, 5\%\}$.]{
  \includegraphics[width=0.3\textwidth]{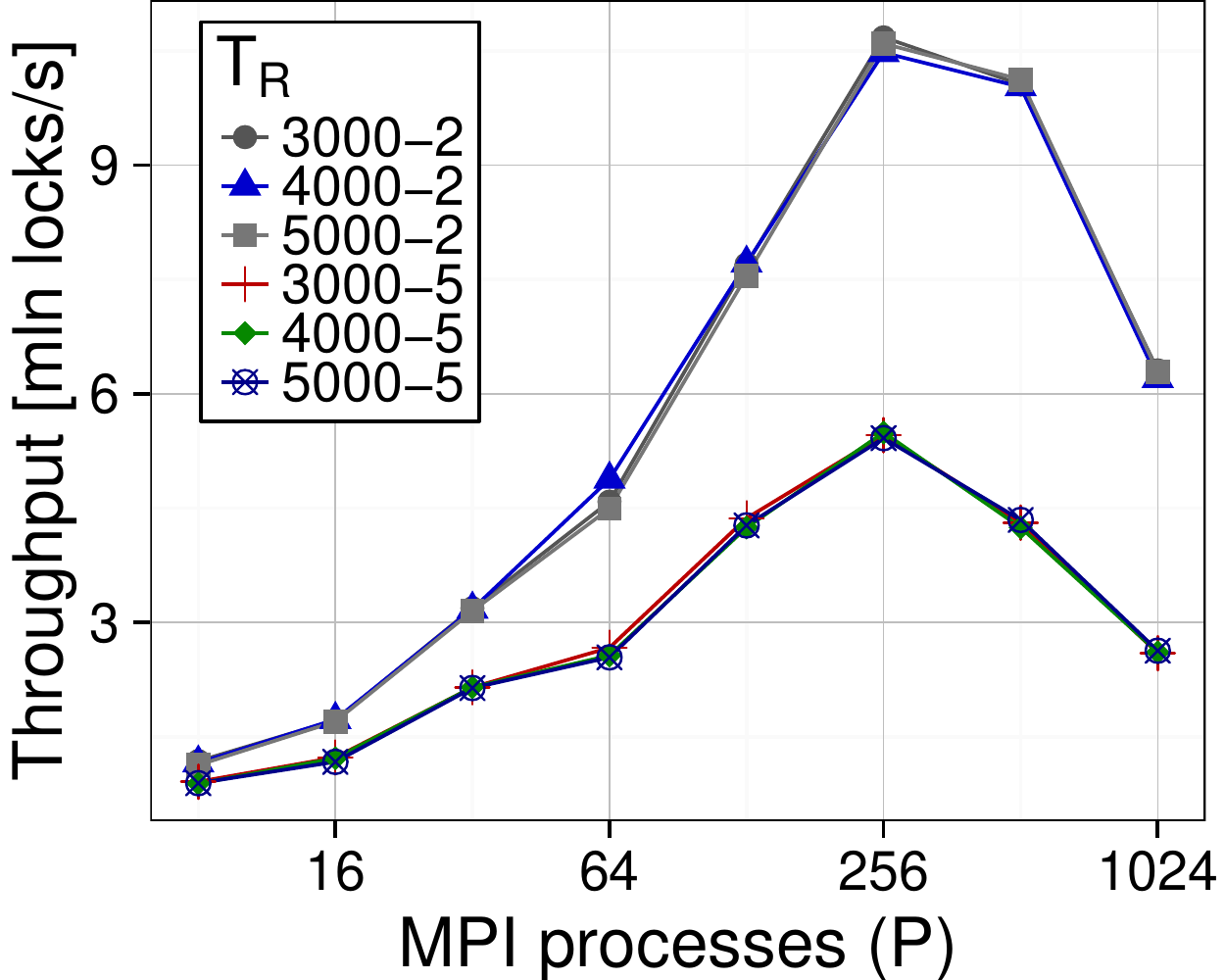}
  \label{fig:rma_rw_T_R_m_rates_ecsb}
 }
 %
%\vspace{-0.5em}
\caption{Analysis of the performance impact of various thresholds.}%\maciej{Add RW rates in each.}}
%\vspace{-1.5em}
\label{fig:rma_rw_thresholds}
\vspace{-0.5em}
\end{figure*}

\textbf{\textsf{Varied Parameters }}
To evaluate various scenarios, we vary: $T_{DC}$, $T_{L,i}$, and $T_{R}$.
Unless stated otherwise, we set the fraction of writers $F_W=0.2\%$ as it
reflects Facebook workloads~\cite{Venkataramani:2012:TFS:2213836.2213957};
however, we also evaluate other values.

% There is an extension for these benchmarks. We use it to optimize the
% thresholds for the writers.  In order to do that, we ensure that these
% thresholds are reached by setting four processes per node as writers and the
% rest as readers. With this feature the percentage of writers is set to 25\%
% on each node. We will refer to this extension by putting a ``W-'' before the
% benchmark names (W-LB for example).

% For all of them, one can choose the percentage of writers respectively
% readers for the RMA-RW and foMPI-RW lock. Before each acquisition, a process
% generates a random number to decide whether it will act as a reader or a
% writer for this iteration. Therefore, there is sometimes an evolution from
% readers to writers and the other way round. 

\textbf{\textsf{Experimentation Methodology }}
To calculate the latency, we derive the arithmetic mean of 100,000
operations per process (for each latency benchmark).  Throughput is the
aggregate count of lock acquires or releases divided by the total time to run a
given benchmark. 10\% of the first measurements are discarded (warmup). All time
measurements are taken using a high precision \texttt{rdtsc}
timer~\cite{hoefler-netgauge-hpcc07}.

\textbf{\textsf{Experimental Setup }}
We conduct experiments on CSCS Piz Daint (Cray XC30). Each node has an 8-core
HT-enabled Intel Xeon E5-2670 CPU with 32 GiB DDR3-1600 RAM. The
interconnection is based on Cray's Aries and it implements the Dragonfly
topology~\cite{faanes2012cray,kim2008technology}. The batch system is slurm
14.03.7. We use C++ and the GNU 5.2.40 g++ compiler with -O3 optimizations. The
utilized Cray DMAPP is 7.0.1-1.0501.8315.8.4.ari.
%
% and an NVIDIA Tesla K20X with 6 GiB GDDR5 RAM
%
Unless stated otherwise, we use all the compute resources and run one MPI
process per one HT resource (16 processes per one compute node).

% All of our experiments were conducted on CSCS Piz Daint (Cray XC30) which
% consists of computing nodes. Each node has an 8-core Intel Xeon E5-2670 CPU
% with 32 GiB DDR3-1600 RAM, an NVIDIA Tesla K20X with 6 GiB GDDR5 RAM. It uses
% GNU's Programming Environment version 5.2.40. Since the cores have the
% technology of hyper-threading, we can have up to 16 processes on one node.
% The interconnection is based on Cray's Aries in a Dragonfly topology. The
% batch system (slurm 14.03.7) chooses the allocated nodes. All our
% implementations were written in C++ and compiled with the GNU 5.2.40 g++
% compiler at optimization level -O3.

\macb{Machine Model }
We consider two levels of the hierarchy: the whole machine and compute
nodes, thus $N=2$.

\textbf{\textsf{Implementation Details }}
We use the
\emph{libtopodisc}~\cite{wgropp} library for discovering the structure of the
underlying compute nodes and for obtaining MPI communicators that enable
communication within each node. We group all the locking structures in MPI
allocated windows to reduce the memory footprint~\cite{fompi-paper}.

% For building up the tree for the hierarchy we need to know how the topology
% looks like. We use \emph{libtopodisc}~\cite{wgropp} for getting a communicator
% between MPI-processes on one computing node. With this communicator a process
% can find out on which node it resides. This is already sufficient for our needs
% since we either communicate within a node or between all MPI-processes. In
% addition, we initialize on each process windows that can be accessed by all
% processes that are associated to the same communicator handle.  

\subsection{Performance Analysis of RMA-MCS}
\label{sec:eval_mcs_lock}

\goal{Explain results of foMPI}
We present the results in Figure~\ref{fig:rma_mcs_perf}.
The latency of RMA-MCS is lower than any other target. For example,
for $P=1,024$, it is $\approx$10x and $\approx$4x lower than \texttt{foMPI-Spin}
and \texttt{D-MCS}, respectively.
This is
because \texttt{foMPI-Spin} entails lock contention that
limits performance. In addition, both \texttt{foMPI-Spin} and
\texttt{D-MCS} are topology-oblivious.
Then, the throughput analysis confirms the advantages of RMA-MCS across all the
considered benchmarks. The interesting spike in ECSB and SOB is because moving
from $P=8$ to $P=16$ does not entail inter-node communication, initially
increasing RMA-MCS's and D-MCS's throughput.
We conclude that RMA-MCS consistently outperforms the original foMPI design 
and D-MCS.

\subsection{Performance Analysis of RMA-RW}
\label{sec:eval_rw_lock}

We now proceed to evaluate RMA-RW. First, we analyze the impact of various
design parameters (Figure~\ref{fig:rma_rw_thresholds}) and then compare it to
the state-of-the-art (Figure~\ref{fig:rma_rw_perf_state-of-the-art}).
Due to space constraints, we only present a subset of the results, all
remaining plots follow similar performance patterns.

% \begin{itemize}
% %
% \item The number of counter in the system
% %
% \item The threshold that regulates after how many consecutive writer lock
% acquisition the lock gets passed to the readers
% %
% \item The thresholds on each level of our tree design that controls the
% conflict between the writers
% %
% \item The threshold for the readers that will make the readers to pass the lock
% to the writers after reaching
% %
% \end{itemize}

\subsubsection{Influence of $T_{DC}$}
\label{sec:eval_t_dc}

We first discuss how different $T_{DC}$ values impact performance. We consider
$T_{DC} \in \{1,2,4\}$ (one physical counter on each compute node and every 2nd and 4th
compute node, respectively). We also vary the number of counters on one
node ($1,2,4,8$).
The results are presented in Figure~\ref{fig:rma_rw_T_DC}. First, lower
$T_{DC}$ entails more work for writers that must access more counters while
changing the lock mode. This limits performance, especially for high $P$,
because of the higher total number of counters. Larger $T_{DC}$ increases
throughput (less work for writers), but at some point (e.g., $P=512$ a counter on
every 2nd node) the overhead due to readers (contention and higher latency) begins
to dominate.
We conclude that selecting the proper $T_{DC}$ is important for high performance
of RMA-RW, but the best value depends on many factors and should be tuned for a
specific machine. For example, higher $T_{DC}$ might entail unpredictable
performance penalties on Cray XE because the job scheduler does not enforce
contiguous job allocations~\cite{bhatele2013there}.

\subsubsection{Influence of $T_{L,i}$}
\label{sec:eval_t_li}

\begin{figure*}
\centering
%\vspace{-1.3em}
 \subfloat[Latency (LB).]{
  \includegraphics[width=0.31\textwidth]{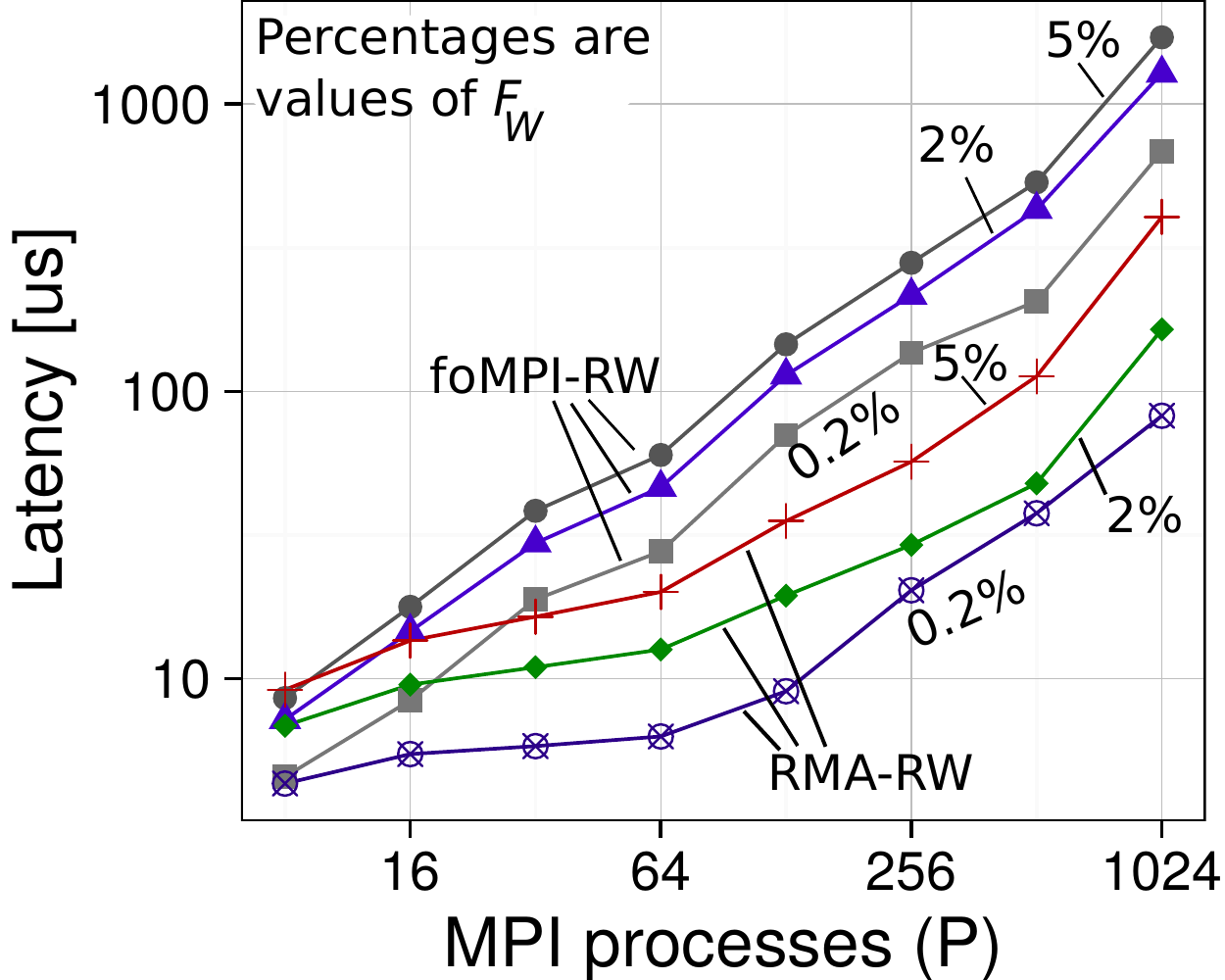}
  \label{fig:s}
 }\hfill
 \subfloat[Throughput (ECSB).]{
  \includegraphics[width=0.31\textwidth]{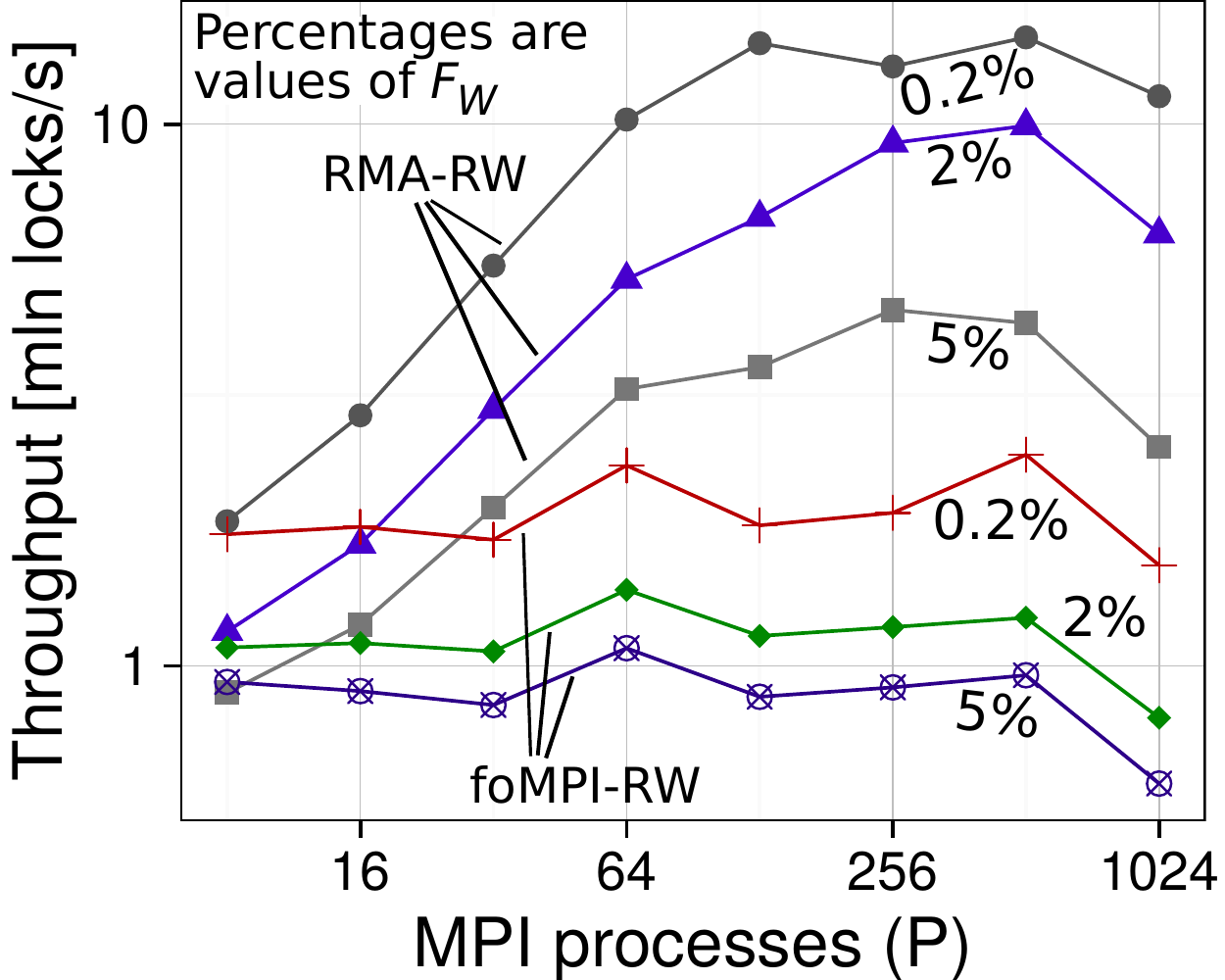}
  \label{fig:q}
 }\hfill
 \subfloat[Throughput (SOB).]{
  \includegraphics[width=0.31\textwidth]{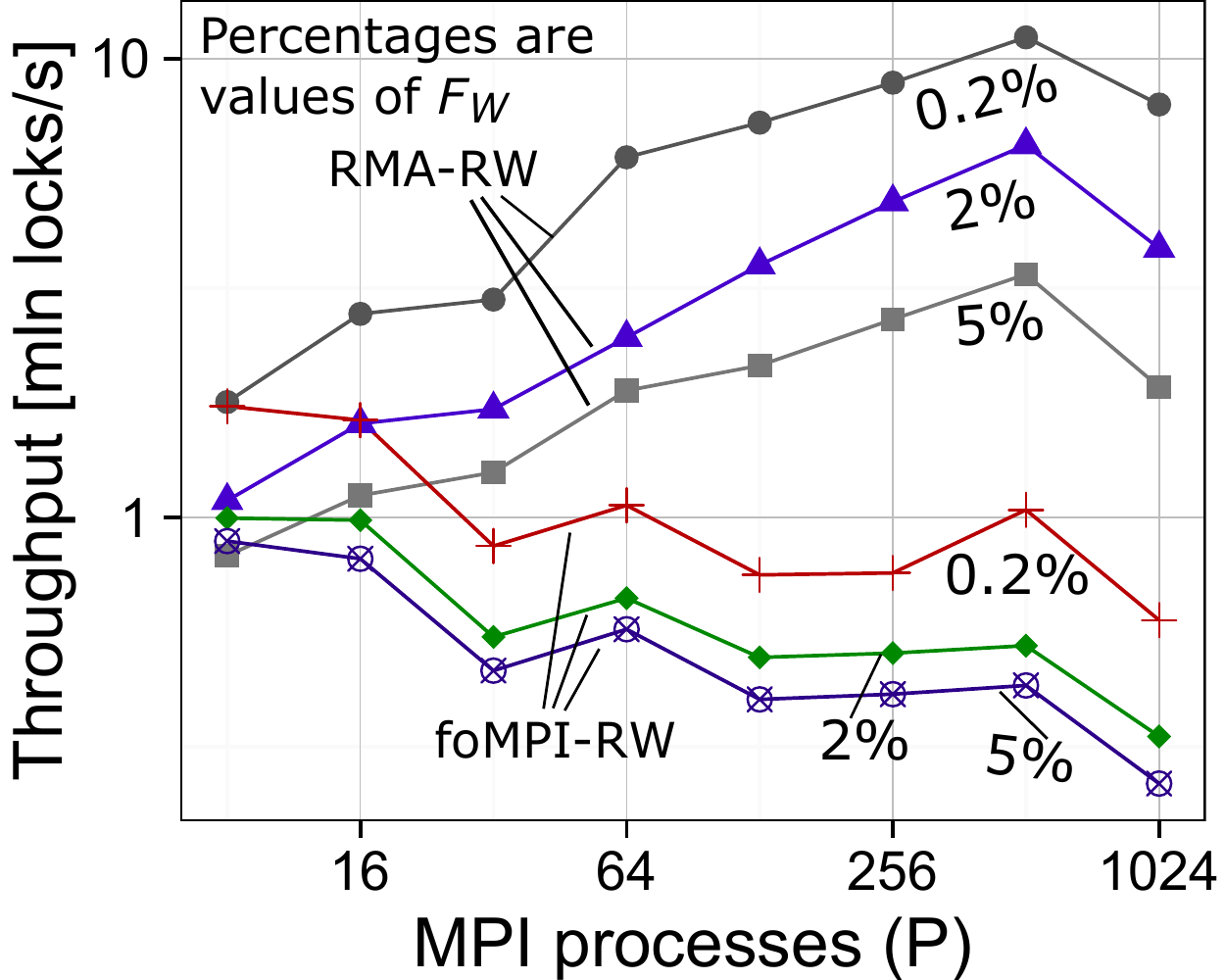}
  \label{fig:q}
 } 
%\vspace{-0.5em}
\caption{(\cref{sec:comp_rw_lock}) Performance analysis of RMA-RW and
comparison to the state-of-the-art.}
%\vspace{-1.5em}
\label{fig:rma_rw_perf_state-of-the-art}
\end{figure*}

\begin{figure*}[t]
\centering
\vspace{-1.3em}
 \subfloat[$F_W = 20\%$.]{
  \includegraphics[width=0.23\textwidth]{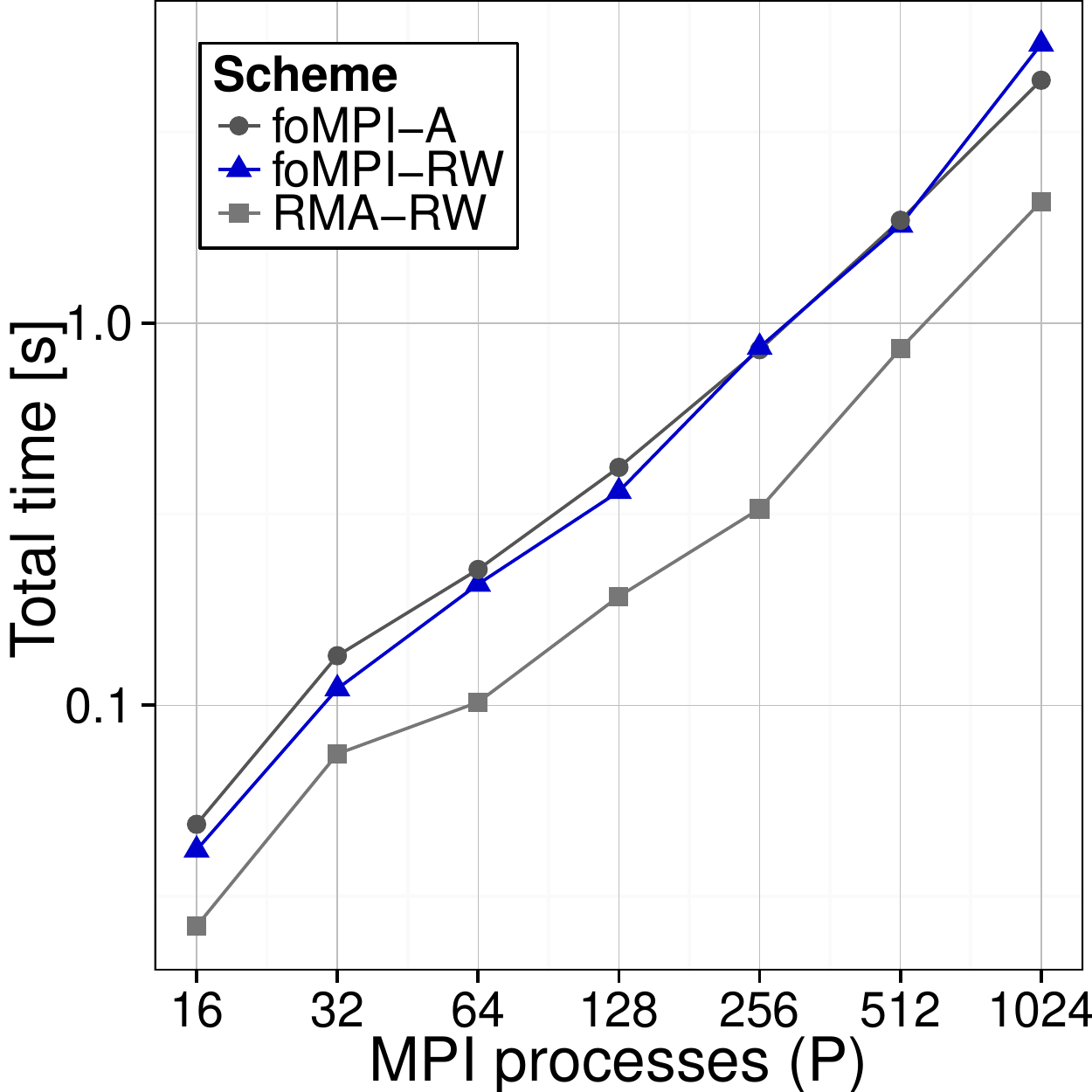}
  \label{fig:ht_rw_80}
 }\hfill
 \subfloat[$F_W = 5\%$.]{
  \includegraphics[width=0.23\textwidth]{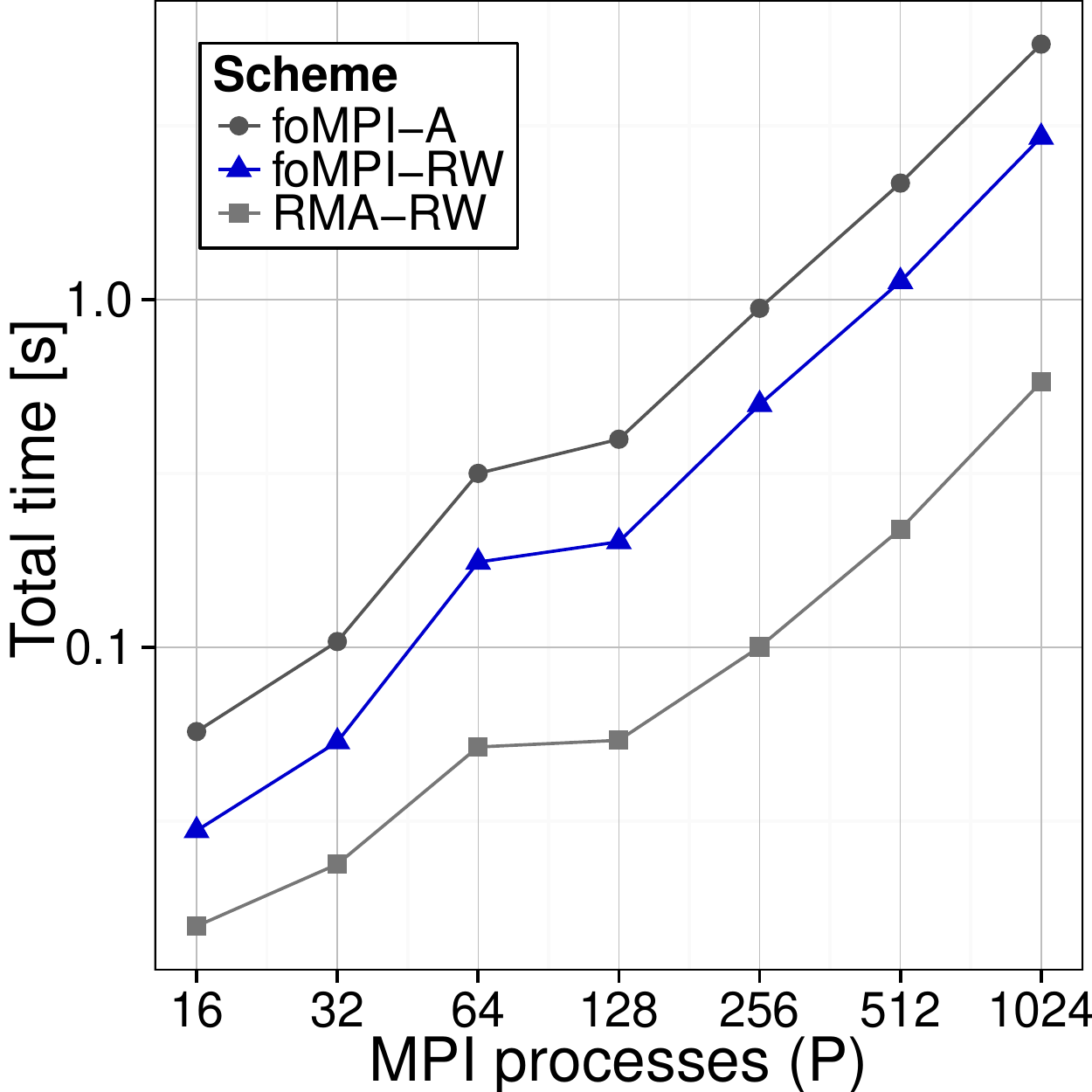}
  \label{fig:ht_rw_95}
 }\hfill
 \subfloat[$F_W = 2\%$.]{
  \includegraphics[width=0.23\textwidth]{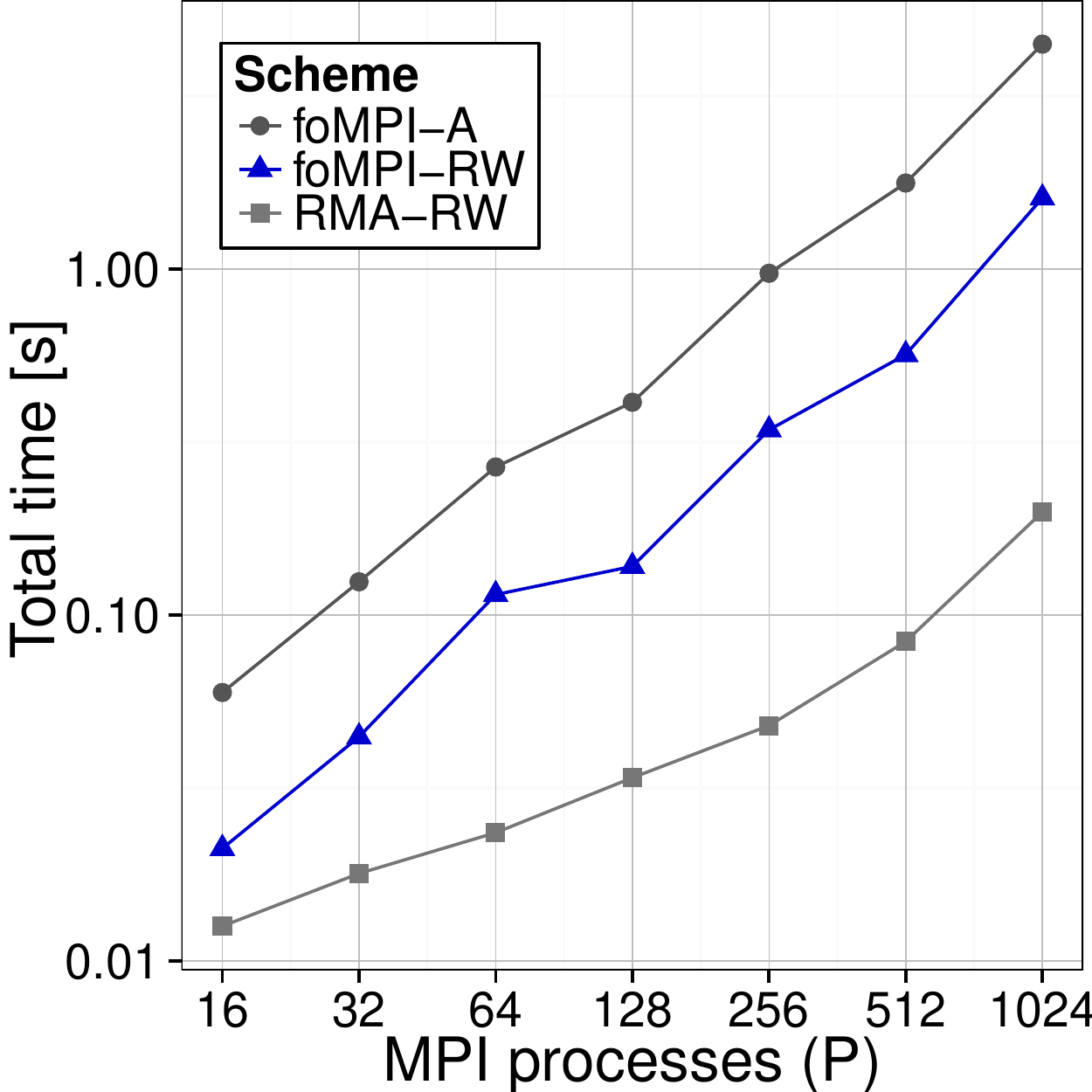}
  \label{fig:ht_rw_98}
 }\hfill
 \subfloat[$F_W = 0\%$.]{
  \includegraphics[width=0.23\textwidth]{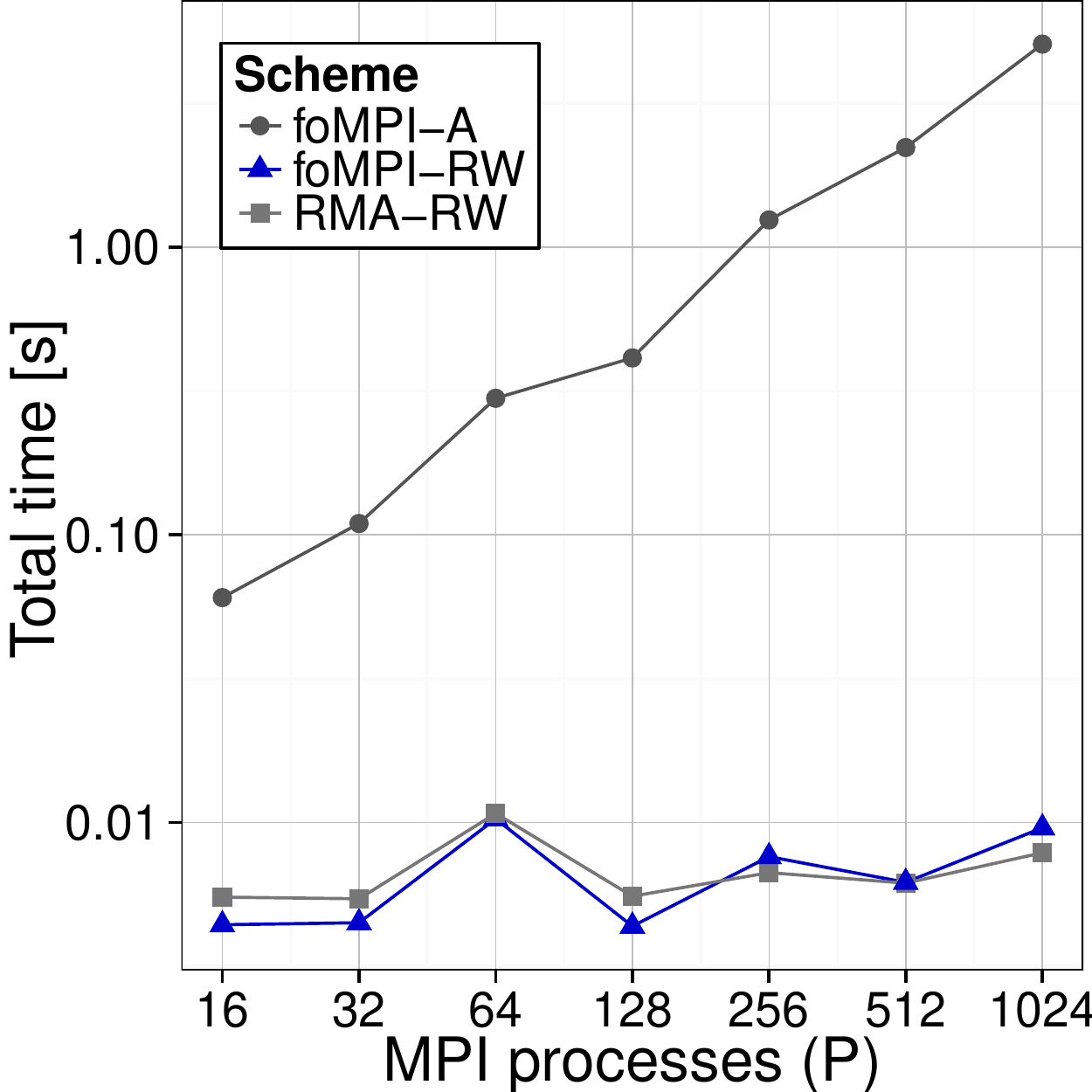}
  \label{fig:ht_rw_100}
 }\hfill
%\vspace{-0.5em}
\caption{(\cref{sec:ht_eval}) Performance analysis of a distributed hashtable.}
%\vspace{-1.5em}
\label{fig:ht_eval}
\vspace{-0.5em}
\end{figure*}

Next, we analyze the performance impact of $T_{L,i}$
in the considered system $i \in \{1,2\}$. We fix $F_W=25\%$ to 
ensure that there are multiple writers per machine element on each level.
We start with various $\prod_{i=1}^{N}T_{L,i}$: the maximal number of writer
acquires before the lock is passed to the readers; see
Figure~\ref{fig:rma_rw_T_Li_prod}. As expected, smaller product increases
throughput because more readers can enter the CS, but reduces fairness as
writers wait longer.
In the second step, we analyze how varying each $T_{L,i}$ impacts performance. 
We first fix $\prod_{i=1}^{N}T_{L,i} = 1000$.
As $N=2$, we 
use $T_{L,2} \in (10,25,50)$ and $T_{L,1} \in (100,40,20)$.
%
% we fix $\prod_{i=1}^{N}T_{L,i} = 1000$
% and then vary each $T_{L,i}$ accordingly. As $N=2$, we 
% use $T_{L,2} \in \{10,25,50\}$ and $T_{L,1} \in \{100,40,20\}$
% and use pairs such that $T_{L,1} T_{L,2} = 1000$.
%
%Such numbers 
%
% want to have a sufficiently large number to exploit the locality and for level 1, we need a value that lets 
% processes on different nodes acquire the lock before handing it to the readers.
%
The outcome is shown in Figure~\ref{fig:rma_rw_T_Li}. When more writers
consecutively acquire the lock within one node (higher $T_{L,2}$), the
throughput increases. Still, the differences between the considered options are
small (up to 25\% of the relative difference), especially for lower $P$.  This
is because of smaller amounts of inter-node communication.
Interestingly, options that increase throughput (e.g., 50-20) also increase
latency, see Figure~\ref{fig:rma_rw_T_Li_LB}.  We conjecture this is due to
improved fairness caused by smaller $T_{L,2}$ (more processes from different
nodes can acquire the lock). However, the average latency increases because
other writers have to wait for a longer time.

\begin{table*}
\centering
\sf
\small
\begin{tabular}{@{}lllllll@{}}
\toprule
 & \textbf{UPC (standard)~\cite{upc}} & \textbf{Berkeley UPC~\cite{bupc}} & \textbf{SHMEM~\cite{shmem}} & \textbf{Fortran 2008~\cite{fortran2008}} & \textbf{Linux RDMA/IB~\cite{ofed-atomics,IBAspec}} & \textbf{iWARP~\cite{iwarp,sharp2014remote}} \\ \midrule
 \texttt{Put} & \texttt{UPC\_SET} & \texttt{bupc\_atomicX\_set\_RS} & \texttt{shmem\_swap} & \texttt{atomic\_define} & \texttt{MskCmpSwap} & masked \texttt{CmpSwap} \\
 \texttt{Get} & \texttt{UPC\_GET} & \texttt{bupc\_atomicX\_read\_RS} & \texttt{shmem\_mswap} & \texttt{atomic\_ref} & \texttt{MskCmpSwap} & masked \texttt{CmpSwap} \\
 \texttt{Accumulate} & \texttt{UPC\_INC} & \texttt{bupc\_atomicX\_fetchadd\_RS} & \texttt{shmem\_fadd} & \texttt{atomic\_add} & \texttt{FetchAdd} & \texttt{FetchAdd} \\
 \texttt{FAO (SUM)} & \texttt{UPC\_INC}, \texttt{UPC\_DEC} & \texttt{bupc\_atomicX\_fetchadd\_RS} & \texttt{shmem\_fadd} & \texttt{atomic\_add} & \texttt{FetchAdd} & \texttt{FetchAdd} \\
 \texttt{FAO (REPLACE)} & \texttt{UPC\_SET} & \texttt{bupc\_atomicX\_swap\_RS} & \texttt{shmem\_swap} & \texttt{atomic\_define}* & \texttt{MskCmpSwap} & masked \texttt{CmpSwap} \\
 \texttt{CAS} & \texttt{UPC\_CSWAP} & \texttt{bupc\_atomicX\_cswap\_RS} & \texttt{shmem\_cswap} & \texttt{atomic\_cas} & \texttt{CmpSwap} & \texttt{CmpSwap} \\ \bottomrule
\end{tabular}
\caption{Illustration of the feasibility of using libraries/languages other than MPI RMA for RMA-MCS/RMA-RW. \texttt{*} indicates the lack of an atomic swap in Fortran 2008,
suggesting that some of RMA-RW protocols that depend on it would have to be adjusted to a different set of available atomics.}
\label{tab:other_libs}
\vspace{-0.5em}
\end{table*}

\subsubsection{Influence of $T_{R}$}
\label{sec:eval_t_rw}

Next, we analyze the impact of $T_{R}$; see
Figure~\ref{fig:rma_rw_T_R_one_rate}. We first use $F_W=0.2\%$. 
The throughput for $T_{R} \in \{$1,000 $;$ 2,000$\}$ drops significantly for $P>512$
due to the higher overhead of writers. Contrarily, increasing $T_{R}$
improves the throughput significantly. This is because the latency of readers is 
lower than that of writers and a higher $T_{R}$ entails a preference of readers. 
However, the larger $T_{R}$ the longer the waiting time for writers is.
%for a level beyond 4,000 (for $P>512$) also diminishes performance. This is
%  because the latency of readers increases for higher $P$ counts as physical
%  distances between nodes may become larger.
%
Finally, we analyze the relationship between $T_{R}$ and $F_W$ in more detail;
%
% see Figures~\ref{fig:rma_rw_T_R_m_rates_LB}-\ref{fig:rma_rw_T_R_m_rates_ecsb}.
%
see Figure~\ref{fig:rma_rw_T_R_m_rates_ecsb}.
Here, we vary $F_W \in \{2\%, 5\%\}$. The results indicate no consistent
significant advantage ($<$1\% of relative difference for most $P$) of one
threshold over others within a fixed $F_W$.

\subsubsection{Comparison to the State-of-the-Art}
\label{sec:comp_rw_lock}

We now present the advantages of RMA-RW over the state-of-the-art foMPI RMA
library~\cite{fompi-paper}; see Figure~\ref{fig:rma_rw_perf_state-of-the-art}.
Here, we consider different $F_W$ rates.
%
% We use the WARB benchmark for fair comparison (WARB scales down the
% contention which improves the results of foMPI). 
%
As expected, any RW distributed lock provides the highest throughput  for $F_W
= 0.2\%$. This is because readers have a lower latency for acquiring a lock
than writers and they can enter the CS in parallel. 
The maximum difference between the rates $F_W = 0.2\%$ and $F_W = 2\%$ is 1.8x
and between $F_W = 0.2\%$ and $F_W = 5\%$ is 4.4x.  We then tested other values
of $F_W$ up to 100\% to find out that for $F_W > 30\%$ the throughput remains
approximately the same.  At such rates, the throughput is dominated by the
overhead of writers that enter the CS consecutively. 

In each case, RMA-RW always outperforms foMPI by $>$6x for $P \ge 64$. One reason for
this advantage is the topology-aware design. Another one is the presence of
$T_{L,i}$ and $T_{R}$ that prevent one type of processes to dominate the other
one resulting in performance penalties.

\subsection{Case Study: A Distributed Hashtable}
\label{sec:ht_eval}

We now illustrate how RMA-RW accelerates a distributed hashtable (DHT) that
represents irregular codes. Our DHT stores 64-bit integers and it consists of
parts called local volumes. Each local volume consists of a table of elements
and an overflow heap for elements after hash collisions. The table and
the heap are constructed with fixed-size arrays. Every local volume is
managed by a different process. 
Inserts are based on atomic CASes. If a collision happens, the losing thread
places the element in the overflow list by atomically incrementing the next
free pointer. In addition, a pointer to the last element is also updated with a
second CAS. Flushes are used to ensure memory consistency.

% To insert an element, a thread atomically updates the pointers to the
% next free cell and the last element in the local volume.

% In the implementation, each process manages
% a part of the hashtable called the local volume consisting
% of a table of elements and an additional overflow heap
% to store elements after collisions. The table and the heap
% are constructed using fixed-size arrays. In order to avoid
% traversing of the arrays, pointers to most recently inserted
% items as well as to the next free cells are stored along with
% the remaining data in each local volume. The elements of
% the hashtable are 8-Byte integers

We illustrate a performance analysis in Figure~\ref{fig:ht_eval}.  In the
benchmark, $P-1$ processes access a local volume of a selected process with a
specified number of inserts and reads targeted at random hashtable elements.
We compare the total execution time of foMPI-A (a variant that only
synchronizes accesses with CAS/FAO), foMPI-RW, and RMA-RW.  For $F_W \in
\{2\%,5\%,20\%\}$ RMA-RW outperforms both the remaining variants.  For $F_w =
0\%$, foMPI-RW and RMA-RW offer comparable performance.

\section{DISCUSSION}
\label{sec:discussion}

\macb{Using Different RMA Libraries/Languages}
In our implementation, we use MPI RMA. Still, the proposed schemes are generic
and can be implemented using several other existing RMA/PGAS
libraries/languages that support the required operations described in
Listing~\ref{lst:rma_calls}. We illustrate this in Table~\ref{tab:other_libs}
(we omit the distinction between blocking and non-blocking operations as any
type can be used in the proposed locks). The analysis indicates that RMA-MCS
and RMA-RW can be used in not only traditional HPC domains (by utilizing UPC,
SHMEM, or RDMA/IB), but also in TCP/IP-based settings (by using iWARP).

\macb{Selecting RMA-RW Parameters}
To set the parameters, we first find an appropriate value for $T_{DC}$. This is
because our performance analysis indicates that $T_{DC}$ has on average the
highest impact on performance of both readers and writers. Here, our evaluation
indicates that placing one counter per compute node results in a reasonable
balance between reader throughput and writer latency. In the second step, we
further influence the reader/writer performance tradeoff by manipulating with
$T_R$ and $T_{L,i}$. To reduce the parameter space, we fix $T_W$ as indicated
in Table~2.  Selecting $T_{L,i}$ depends on the hardware hierarchy and would
ideally incorporate several performance tests before fixing final numbers. One
rule of the thumb is to reserve larger values for $T_{L,i}$ associated with
components with higher inter-component communication costs, such as racks; this
may reduce fairness, but increases throughput.
%
% Our evaluation indicates that the above strategy is enough for high performance
% and exhaustive searches for optimal values of the parameters are not required.
% Our strategy always resulted in performance higher than any other considered
% distributed lock. We will include these recommendations in the final
% submission.

% \begin{description}
% %
% \item[UPC~\cite{upc}:] the required atomic operations may come
% either from the UPC standard, or the Berkeley UPC extension.
% In the former, we have the following macros: \texttt{UPC\_SET}
% {REPLACE}), 
% \texttt{UPC\_INC} and \texttt{UPC\_SUB} (\texttt{FAO}/\texttt{Accumulate} and \texttt{SUM}),
% \texttt{UPC\_CSWAP}
% (\texttt{CAS}).
% %
% The latter offers \texttt{bupc\_atomicX\_}
% %
% \end{description}

\section{RELATED WORK}

\textbf{\textsf{Queue-Based Locks}}
The well-known traditional examples of this family are
CLH~\cite{Craig93buildingfifo, Magnusson:1994:QLC} and
MCS~\cite{Mellor-Crummey:1991:ASS}.  Yet, they are
oblivious to the memory hierarchy and cannot use this knowledge 
to gain performance.
More recently, Radovic and Hagersten~\cite{Radovic:2003:HBL} proposed a
hierarchical backoff lock that exploits memory locality: a thread reduces its
backoff delay if another thread from the same cluster owns the lock. This
increases the chance to keep the lock within the cluster, but introduces the
risk of starvation. Luchangco et al.~\cite{luchangco2006hclh} improved this
scheme by introducing a NUMA-aware CLH queue that ensures no starvation. Yet,
it considers only two levels of the memory hierarchy. Chabbi et
al.~\cite{Chabbi:2015:HPL:2688500.2688503} generalized it to any number of
memory hierarchy levels. Similarly to our scheme, they introduce an MCS lock
for each level. Yet, they do not target DM machines. None of these protocols
  can utilize the parallelism of miscellaneous workloads where the majority of
  processes only read the data.

\noindent
\textbf{\textsf{RW Locks}}
There exist various traditional RW proposals~\cite{Hsieh:1992:PPS,
Krieger93afair}.  Recently, Courtois et
al.~\cite{Courtois:1971:CCL} introduced different preference schemes that favor
either readers (a reader can enter the CS even if there is a writer waiting) or
writers (a writer can enter the CS before waiting readers).  Yet, this protocol
neither prevents starvation nor scales well. Mellor-Crummey and
Scott~\cite{Mellor-Crummey:1991:SRS} extended their MCS lock to distinguish
between readers and writers. This algorithm however does not scale well under
heavy read contention. Next, Krieger et al.~\cite{Krieger93afair}
use a double-linked list for more flexibility in how processes traverse the
queue. Yet, there is still a single point of contention. Hsieh and Weihl
\cite{Hsieh:1992:PPS} overcome this by trading writer throughput for reader
throughput. In their design, each thread has a private mutex; the readers
acquire the lock by acquiring their private mutex but the writers need to
obtain all mutex objects.  This introduces a massive overhead for the writers
for large thread counts.
Other approaches incorporate elaborate data structures like the Scalable
Non-Zero Indicator (SNZI) tree~\cite{Lev:2009:SRL} that traces readers in the
underlying NUMA hierarchy for more locality. Yet, writers
remain NUMA-oblivious. Calciu et al.~\cite{Calciu:2013:NRL} extend this
approach with an RW lock in which both readers and writers are NUMA-aware. This
design improves memory locality but it only considers two levels in a NUMA
hierarchy.
None of these schemes address DM environments.

\noindent
\textbf{\textsf{Distributed Locks}}
To the best of our knowledge, little research has been performed into locks for
DM systems. Simple spin-lock protocols for implementing MPI-3 RMA
synchronization were proposed by Gerstenberger et al.~\cite{fompi-paper}. Some
other RMA languages and libraries (e.g., UPC) also offer locks, but they are
not RW, their performance is similar to that of foMPI, and they are
hardware-oblivious.

We conclude that our work offers the first lock for
DM systems that exploits the underlying inter-node structure
and utilizes the RW parallelism present in various data- and
communication-intensive workloads.

\section{CONCLUSION}

Large amounts of data in domains such as graph computations require
distributed-memory machines for efficient processing. Such machines are
characterized by weak memory models and expensive inter-node communication.
These features impact the performance of topology-oblivious
locks or completely prevent a straightforward adoption of
existing locking schemes for shared-memory systems.

In this work, we propose a distributed topology-aware Reader-Writer (RMA-RW)
and MCS lock that outperform the
state-of-the-art. RMA-RW offers a modular design with three parameters that
offer performance tradeoffs in selected parts of the lock. These are: higher
lock fairness or better locality, larger throughput of readers or writers, and
lower latency of readers or writers. This facilitates performance tuning for a
specific workload or environment. RMA-RW could also be extended with adaptive
schemes for a runtime selection and tuning of the values of the parameters.
This might be used in accelerating dynamic workloads.

Microbenchmark results indicate that the proposed locks
outperform the state-of-the-art in both latency and throughput. 
Finally, RMA-RW accelerates a distributed hashtable that represents
irregular workloads such as key-value stores.

% These numbers show that there is a large potential in topology-aware lock
% algorithms. They exploit the locality of data which makes quite a difference
% regarding the latency of communication. Therefore, we encourage researchers
% to find synchronization structures and concurrent algorithms that utilize
% this potential. 

% \maciej{Patrick: when writing tests for correctness, did you use N>2 ?}
% 
% \maciej{Make counters sit on two cache lines}
% 
% \maciej{check the goals (add to the checklist, + all others, and check the checklist}
% 
% \maciej{Make thresholds dynamic?}
% 
% \maciej{I removed (commented) some stuff that seemed unnecessary, e.g., parent/next pointer to the up/next DQ, PARENT offset, etc. Double check.}
% 
% \maciej{Add the part the reset counters: 
% ``We reset the counters every time
% $T_{R}$ is reached (to prevent counter overflows) or when no reader is in the
% CS but writers are waiting (to prevent starvation of writers). ''}
% 
% 
% \maciej{General remark - we may want to rerun some stuff, are these any scripts
% that facilitate collecting results?}

{\section*{Acknowledgements} 
This work was supported by Microsoft
Research through its Swiss Joint Research Centre.
We thank our shepherd Patrick G.~Bridges, anonymous reviewers, and Jeff Hammond for their insightful comments.
We thank the CSCS
team granting access to the Piz Dora and Daint machines, and for
their excellent technical support.}

\bibliographystyle{abbrv}
\bibliography{references}

%\balancecolumns
%\appendix

\end{document}